\documentclass[a4paper,11pt]{article}
\pdfoutput=1
\usepackage{jheppub}
\usepackage[T1]{fontenc}
\usepackage{subfigure}
\usepackage{paralist}
\usepackage{booktabs}
\usepackage{multirow}
\usepackage{float}
\usepackage{placeins}
\restylefloat{table}
\title{\boldmath Non-parametric data-driven background modelling using conditional probabilities}
\author[]{Andrew~Chisholm,}
\author[]{Thomas~Neep,}
\author[]{Konstantinos~Nikolopoulos,}
\author[1]{Rhys~Owen,\note{Now at Rutherford Appleton Laboratory, Science and Technology Facilities Council.}}
\author[2]{Elliot~Reynolds\note{Corresponding author. Now at Physics Division, Lawrence Berkeley National Laboratory.}}
\author[]{and J\'ulia~Silva}
\affiliation[]{School of Physics and Astronomy, University of Birmingham,\\Birmingham, B15 2TT, United Kingdom}
\emailAdd{andrew.chisholm@cern.ch}
\emailAdd{tom.neep@cern.ch}
\emailAdd{konstantinos.nikolopoulos@cern.ch}
\emailAdd{rhys.owen@cern.ch}
\emailAdd{elliot.reynolds@cern.ch}
\emailAdd{julia.manuela.silva@cern.ch}

\abstract{Background modelling is one of the main challenges in particle physics data analysis. Commonly employed strategies include the use of simulated events of the background processes, and the fitting of parametric background models to the observed data. However, reliable simulations are not always available or may be extremely costly to produce. As a result, in many cases uncertainties associated with the accuracy or sample size of the simulation are the limiting factor in the analysis sensitivity. At the same time, parametric models are limited by the a priori unknown functional form and parameter values of the background distribution.
These issues become ever more pressing when large datasets become available, as it is already the case at the CERN Large Hadron Collider, and when studying exclusive signatures involving hadronic backgrounds.

A widely applicable approach for non-parametric data-driven background modelling is proposed, which addresses these issues for a broad class of searches and measurements. It relies on a relaxed version of the event selection to estimate conditional probability density functions and two different techniques are discussed for its realisation.
The first relies on ancestral sampling and uses data from a relaxed event selection to estimate a graph of conditional probability density functions of the variables used in the analysis, while accounting for significant correlations. A background model is then generated from events drawn from this graph, on which the full event selection is applied.
In the second, a novel generative adversarial network is trained to estimate the joint probability density function of the variables used in the analysis. The training is performed on a relaxed event selection, which excludes the signal region, and the network is conditioned on a blinding variable.
Subsequently, the conditional probability density function is interpolated into the signal region to model the background.
The application of each method on a benchmark analysis and on ensemble tests is presented in detail, and the performance is discussed.}

\usepackage{xspace}

\newcommand*{\GeV}{\ensuremath{\text{Ge\kern -0.1em V}}}
\newcommand*{\MeV}{\ensuremath{\text{Me\kern -0.1em V}}}
\newcommand*{\gev}{\ensuremath{\text{Ge\kern -0.1em V}}}
\newcommand*{\pt}{\ensuremath{p_{\text{T}}}\xspace}

\begin{document}
\maketitle

\section{Introduction}
\label{sec:intro}

The modelling of background processes is a critical element in
determining the discovery potential of hadron collider experiments
searching for physics beyond the Standard Model (SM).  In this
direction, data-driven background estimation methods have been
deployed.  In simple cases, it is sufficient to estimate the expected
number of background events in a given signal region, for example
through single or double sideband methods. But in most cases, reliable
description of the background shape in a discriminant variable is also
required.  For these cases, background modelling often relies on
direct simulation based on Monte Carlo (MC) event generators or on
parametric methods.

In many physics analyses, however, the dominant contribution to the
total background cannot be modelled with sufficient accuracy using MC
simulations.  Typical examples include fully hadronic final states and
backgrounds associated with the mis-identification of physics objects
at the reconstruction level.  Furthermore, the composition of the
background itself, in terms of distinct scattering processes, may not
be reliably known. This can lead to a situation where the
characteristics of the total background cannot be accurately
predicted, even if some of the known components can be modelled with
sufficient accuracy.

In the case of parametric methods, the shape of the background in the
variable of interest is interpolated, or less often extrapolated,
using a functional form modelling the underlying background
distribution.  The parameters of the function are obtained directly
from a fit to the data.  Challenges associated with this approach
include identifying the optimal functional form: a function with too
many free parameters will result in a reduced statistical power, while
one with too few parameters may not have enough flexibility to
describe the underlying distribution.  Fundamentally, however, there
is no guarantee that the actual distribution of the background is part
of the family of curves parameterised by the chosen function. This may
directly bias the extraction of a potential signal.  Several
approaches have been devised to quantify the potential bias associated
with this issue. One such approach is to perform ``spurious-signal''
calculations~\cite{ATLAS:2012yve,ATLAS:2020fzp}, which rely on
computationally intensive MC simulations. Another approach involves
discrete profiling of an ensemble of parametric
forms~\cite{Dauncey:2014xga,CMS:2020xrn,CMS:2020xwi}, including a
penalty factor to account for differences in the number of free
parameters.  This penalty factor needs, in principle, to be determined
in a case-by-case basis, considering coverage properties and any
residual bias. A further complication arises in the presence of
correlated uncertainties across different event categories, where
although the application of the method is conceptually simple, in
practice approximations are required.  It is noted, that the modeling
of smooth backgrounds with non-parametric methods has also been
proposed, for example through Gaussian Processes~\cite{Frate:2017mai}.

In this article, a widely applicable approach for non-parametric
data-driven background modelling is proposed to derive a reliable
understanding of the background to searches for physics beyond the SM
in decays of the Higgs boson. It is general enough to be applicable to
a broad spectrum of analyses, including both searches and precision
measurements, and utilises directly a relaxed version of the event
selection to estimate conditional probability density functions. Two
different techniques are discussed for its realisation: the first,
based on the concept of ancestral sampling, is demonstrated within the
context of a search for rare exclusive hadronic decays of the Higgs
boson~\cite{Aaboud:2017xnb,Aaboud:2018txb,Aaboud:2016rug,Aad:2015sda},
while the second, based on novel generative adversarial networks, is
demonstrated within the context of a search for Higgs boson decays to
a $Z$ boson and an undiscovered light resonance, decaying to a
low-multiplicity hadronic final state~\cite{Aad:2020hzm}.

The two case studies presented here are analysed using separate
samples of simulated signal and background events. The event generator
configuration specific to the $H\to\phi\gamma$ search is described in
section~\ref{analysis_phiy} while the corresponding configuration for
the search for $H\to Za\to \mu\mu + {\rm jet}$ is described in
section~\ref{sec:cgan_event}.  For both channels, generated events are
processed with the Delphes fast simulation framework (version
3.4.2)~\cite{deFavereau:2013fsa}. This framework uses parameterised
descriptions of the response of collider experiment detectors to
provide reconstructed physics objects, enabling a realistic data
analysis to be performed. As an example of a general purpose Large
Hadron Collider (LHC) detector, the ATLAS-like configuration card
included in Delphes is used. This is minimally modified to use the
charged hadron tracking resolution from the earlier version 3.3.3,
which was found to more closely reproduce the performance of the LHC
general purpose detectors~\cite{CMS:2014pgm}.  The effect of multiple
$pp$ interactions in the same or neighbouring bunch crossings (pileup)
are not simulated, since it represents a negligible contribution to
the background in both case studies.

In both implementations, a generative model is derived that aims to
reproduce the kinematic and related event selection variables that are
subsequently used to model a compound variable of interest, for
example the invariant mass of the system. Conditional probabilities
have also been employed for background parameterisation through
density estimation~\cite{Mathad:2019rqj,Nachman:2020lpy}. Typically,
the density estimate is obtained from the sidebands and is
interpolated (or extrapolated) to the signal region, while it is made
conditional to one or more variables of interest. Furthermore, in an
earlier work~\cite{Andreassen:2020nkr} the ``deep neural networks
using classification for tuning and reweighting''
procedure~\cite{Andreassen:2019nnm} was used to reweight a background
simulation in the sideband of a resonant feature of the potential
signal. The training of the neural network was conditional on that
feature variable, which enables the subsequent interpolation to the
signal region.

\section{Background modelling with ancestral sampling}
\label{sec:npdd}
Consider a hypothetical particle $X$, decaying to a multi-body final
state.  In order to establish the existence of such a hypothetical
particle within a dataset, one typically begins by identifying
candidate decay products within an event and studying the distribution
of compound kinematic variables of the reconstructed system.  When
studying the distributions of such compound kinematic variables,
typically the invariant mass, for evidence of a new particle, an
accurate description of these distributions for the background is a
primary ingredient in the statistical analysis and interpretation of
the observed events.

The method outlined in this article, based on the concept of ancestral
sampling~\cite{Goodfellow-et-al-2016}, centres around a construct
termed a ``kernel''. The kernel represents a description of the
distributions and correlations among a set of variables, which include
the components of the four vectors, possibly in addition to other
quantities, such as an isolation measure, of each of the reconstructed
decay products.  From this kernel, a data structure termed a
``pseudo-candidate''\footnote{The term ``pseudo-candidate'' denotes a
data structure containing a four vector for each of the final state
particles and possibly further quantities, such as isolation
measures.} can be randomly sampled, such that an ensemble of
pseudo-candidates will respect both the distributions of, and
correlations among, the same variables in the data sample used to
build the kernel.

In the simplest case of the two-body decay $X\to a b$, where only the
four vectors of decay products $a$ and $b$ are described, the most
general form of the kernel is an eight dimensional distribution of the
four vector components of the two decay products. Given that the
method is designed to be built from a dataset, such as a sample of
data events collected by a collider experiment, the kernel may be an
eight dimensional histogram built from the source dataset.

Histograms of dimensionality in excess of three are often not
practical.  It is therefore necessary, in order to limit the impact of
statistical fluctuations, to minimise the dimensionality of the
kernel.  It is often possible to substantially reduce the
dimensionality of the kernel through a judicious choice of the
variables used and a good understanding of the correlations among
them.  An ideal choice is one that minimises the number of strong
correlations among variables.  In such a case, the description of the
kernel can be factorised from a single distribution of high
dimensionality into multiple distributions of low dimensionality. In
this way, only the most important correlations among variables are
explicitly described, while the remaining small correlations are
effectively removed.  The extent of this factorisation can be chosen
based on the accuracy required and the size of the dataset available
to build the kernel.

Once the kernel has been established, an ensemble of pseudo-candidates
can be generated. Each pseudo-candidate is built by randomly sampling
values from each of the factorised distributions in a sequential
manner.  Compound kinematic quantities for each pseudo-candidate, such
as the invariant mass or transverse momentum of the system of two
decay products, can be calculated on a per-candidate basis; this forms
the basis upon which a model describing the distribution of compound
kinematic quantities is derived.

{\boldmath\subsection{Overview of case study: search for $H\to\phi(K^{+}K^{-})\gamma$ }}
The coupling of the Higgs boson to the first and second generation
fermions is yet to be confirmed experimentally.  Rare exclusive decays
of the Higgs boson into a light meson and a photon have been suggested
as a probe of the couplings of the Higgs boson to light
quarks. Specifically, the Higgs boson decay into a $\phi$ meson, where
the decay $\phi\to K^{+}K^{-}$ is considered, and a photon is
sensitive to the Higgs boson coupling to the strange quark.  The ATLAS
collaboration has performed a search for such decays using a dataset
corresponding to $36\,\text{fb}^{-1}$ of $\sqrt{s}=13\,\text{TeV}$
$pp$ collisions\,\cite{Aaboud:2017xnb}.  The final state consists of a
pair of oppositely charged kaons, with an invariant mass consistent
with the $\phi$ meson mass, recoiling against an isolated photon. The
main sources of background in this search are events involving
inclusive direct photon production or multijet processes where a meson
candidate is reconstructed from charged particles produced in a
hadronic jet. Such processes are difficult to model accurately with MC
event generators and represent an ideal use case for a data-driven
non-parametric background modelling method.  The ancestral
sampling-based technique described in this paper has been successfully
deployed in several ATLAS searches for radiative Higgs boson decays to
light
mesons~\cite{Aaboud:2017xnb,Aaboud:2018txb,Aaboud:2016rug,Aad:2015sda}.
In this paper the original technique and a modified implementation are
presented, their statistical properties are characterised, and their
relationship is discussed.

\subsection{Event selection, analysis strategy and simulation}
\label{analysis_phiy}
The event selection used in this case study closely follows that
described in ref.~\cite{Aaboud:2017xnb}.  Events are required to
contain a photon with transverse momentum, $\pt(\gamma)$, in excess of
$35\,\GeV$, and pseudorapidity, $\eta(\gamma)$, within $|\eta(\gamma)|
< 2.37$. Photon candidates with pseudorapidity in the transition
region between barrel and endcap calorimeter regions, $1.37<
|\eta(\gamma)|<1.52$, are excluded. Candidate $\phi\to K^{+}K^{-}$
decays are reconstructed from pairs of oppositely charged tracks with
transverse momentum, $\pt(K)$, in excess of $15\,\GeV$, and absolute
pseudorapidity, $|\eta(K)|$, less than $2.5$.  Furthermore, the
highest transverse momentum track within a pair is required to satisfy
$\pt(K)>20\,\GeV$. These $\pt$ requirements reflect standard trigger
thresholds employed in searches for such decay topologies.  The
invariant mass of track pairs is required to be within the range
$1.012 < m(\phi) < 1.028\,\GeV$, which accounts for the resolution of
the detector. The $H\to\phi\gamma$ candidate is formed from the
combination of the photon with the highest transverse momentum and the
$\phi\to K^{+}K^{-}$ candidate with an invariant mass closest to the
$\phi$ meson mass.  The variable $I(\phi)$ characterises the hadronic
isolation of the $\phi\to K^{+}K^{-}$ candidates. $I(\phi)$ is defined
as the scalar sum of the $\pt$ of tracks within $\Delta R =
\sqrt{\Delta\Phi^2+\Delta\eta^2}=0.2$, where $\Phi$ is the azimuthal
angle, of the leading track within a $\phi\to K^{+}K^{-}$ meson
candidate, relative to the transverse momentum of the $\phi\to
K^{+}K^{-}$ candidate. The tracks constituting the $\phi$ meson
candidate are excluded from the sum.  Events are retained for further
analysis if $\Delta \Phi (\phi, \gamma) > \pi/2$ and $I(\phi) < 0.5$,
targeting isolated $\phi$ meson candidates and decay products that are
created back-to-back. The transverse momentum of the $\phi$ meson
candidate is required to be greater than a threshold that varies as a
function of the invariant mass $m_{\phi\gamma}$ of the three-body
system. Thresholds of $40\, \GeV$ and $47.2\, \GeV$ are imposed on
$\pt(\phi)$ for the regions $m_{\phi\gamma} < 91\, \GeV$ and
$m_{\phi\gamma} \geq 140\, \GeV$, respectively. The threshold is
varied from $40\, \GeV$ to $47.2\, \GeV$ as a linear function of
$m_{\phi\gamma}$ in the region $91 \leq m_{\phi\gamma} < 140\,
\GeV$. This selection was optimised for a simultaneous search for
Higgs and Z bosons decaying to $\phi\,\gamma$.  This set of criteria
define the ``Signal Region'' (SR) requirements.  $H\to\phi\gamma$
signal events are discriminated from background events by means of a
statistical analysis of the distribution of the invariant mass
$m(\phi,\gamma)$ of selected candidates.

For the purpose of demonstrating the ancestral sampling background
modelling method, only the dominant contributions to the signal and
background processes are explicitly simulated. This choice has no
implications for the validation of the method and is pragmatically
motivated. Inclusive Higgs boson production in $pp$ collisions is
approximated by the gluon-fusion process alone and simulated with the
Pythia 8.244 MC event generator~\cite{Sjostrand:2014zea} with the
CT14nlo PDF set~\cite{Dulat:2015mca}. Subleading contributions from
the vector boson fusion process and Higgs boson production in
association with vector bosons and heavy quarks are not simulated
explicitly.  The $H\to\phi\,\gamma$ decay is simulated directly by the
Pythia 8.244 MC event generator~\cite{Sjostrand:2014zea} and no other
Higgs boson decays are simulated.  The $\gamma+\text{jet}$ process
alone is used as a proxy for the inclusive background to the
$H\to\phi\gamma$ search, which is expected to also contain
contributions from multijet events. The production of
$\gamma+\text{jet}$ is simulated with the Sherpa 2.2.10 event
generator~\cite{Gleisberg:2008ta} with the NNPDF3.0 PDF
set~\cite{NNPDF:2014otw}.  Direct photon production with up to two
additional jets is simulated at the matrix element level.  The
simulated $\gamma+\text{jet}$ sample contains around $3\times10^8$
unweighted events and corresponds to an effective integrated
luminosity of around $20\,\text{fb}^{-1}$.

\subsection{Overview of method}
The procedure is based on a sample of data events selected based on
the nominal ``Signal Region'' requirements described in
section~\ref{analysis_phiy}, modified by relaxing a number of criteria
in order to enrich the sample in background events. The criteria which
define this background-dominated sample are denoted the ``Generation
Region'' (GR).  Two additional event samples, known as ``Validation
Regions'' (VR), are also defined to validate the background modelling
procedure. The VR event samples are non mutually exclusive subsets of
the GR sample and the definitions of these regions are outlined in
table~\ref{table:regions}.

\begin{table}[!h]
\centering
\begin{tabular}{|c|c|c|}
  \hline
        & Minimum $\pt(\phi)$ requirement & Maximum $I(\phi)$ requirement \\
        \hline
    GR & $35\,\GeV$ & Not applied \\
    VR1 & Varying from 40 to $47.2\,\GeV$  & Not applied \\
    VR2 & $35\,\GeV$ & 0.5 \\
    SR & Varying from  40 to $47.2\,\GeV$ & 0.5 \\
      \hline
\end{tabular}
\caption{The event selection criteria of the ``Signal Region'' which are modified to define the "Generation Region" and "Validation Regions".\label{table:regions}}
\end{table}

These $n$ events in the GR are used to construct probability density
functions of the relevant kinematic and isolation variables,
parameterised to respect the most important correlations.  This can be
achieved either by directly utilising each of the $n$ events once, as
in
refs.~\cite{Aaboud:2017xnb,Aaboud:2018txb,Aaboud:2016rug,Aad:2015sda},
or by sampling $n$ events from the GR dataset with replacement.  These
approaches are found to produce background models of equal accuracy,
as outlined in section~\ref{sec:NPDD_toys}. In the case study that
follows, results from the former are presented and comparisons to the
latter are made where relevant differences occur.

By sampling the obtained distributions, an ensemble of
$H\to\phi\gamma$ pseudo-candidates is generated, from which a model of
the $m(\phi,\gamma)$ distribution for background events can be derived
for the SR requirements.  In addition to providing a prediction for
the shape of the $m(\phi,\gamma)$ distribution, the normalisation of
the SR and VRs, relative to the GR, is also predicted. The absolute
normalisation in the SR and VRs may be obtained directly by the model,
by uniformly scaling the distributions by the ratio of the number of
data events in the GR to the size of the ensemble of pseudocandidates.

\subsection{Sampling procedure}
\begin{figure}[!h]
  \centering
  \includegraphics[width=0.55\textwidth]{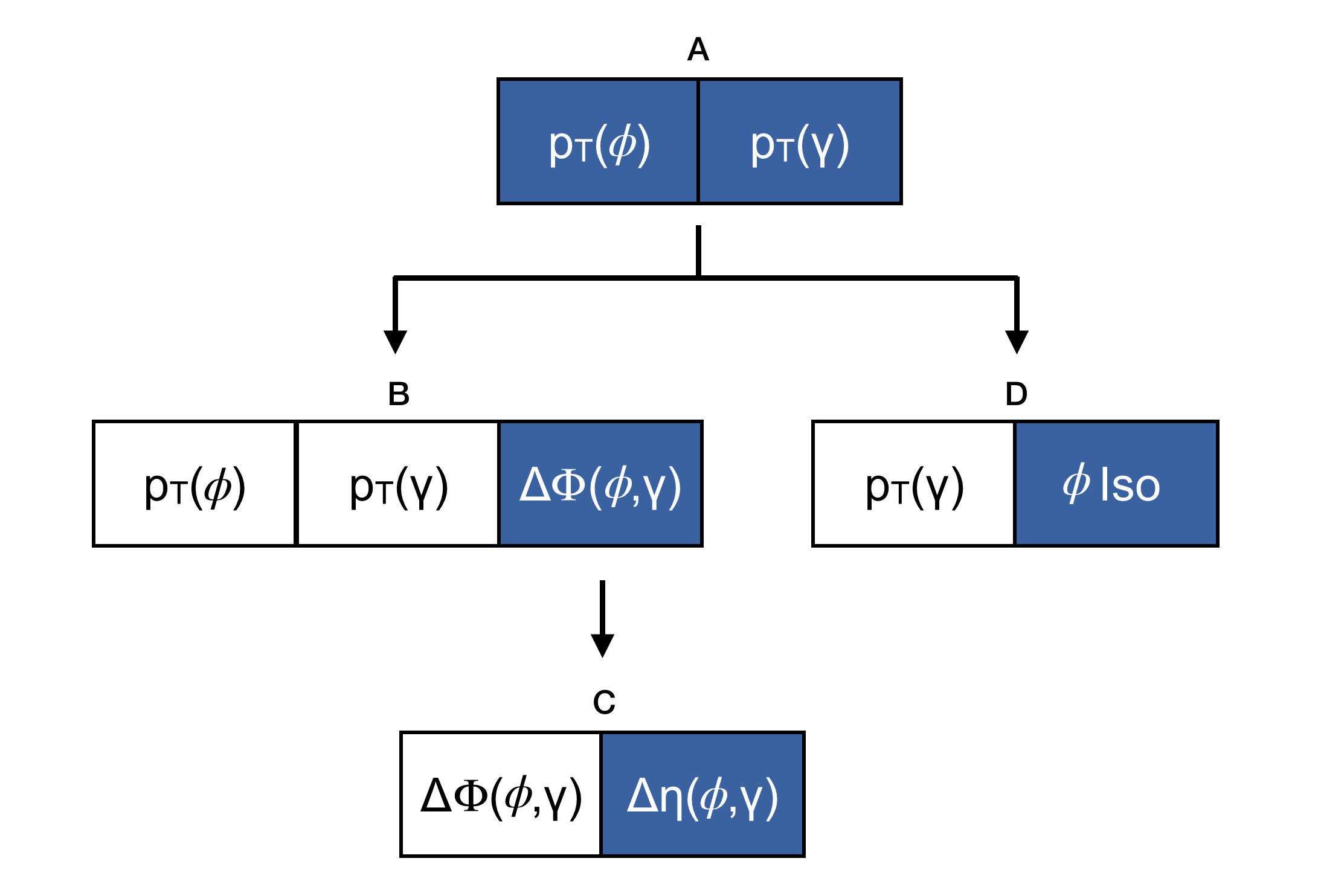}
  \caption{Graphical representation of the sampling sequence followed in the modelling. Variables not shown explicitly are sampled in a factorised, uncorrelated, manner from an 1D template, as described in table\,\ref{table:templates}.}
  \label{fig:sampling}
\end{figure}

Each pseudo-candidate event is described by $\phi$ and $\gamma$
four-momentum vectors and an associated $\phi$ hadronic isolation
variable, $I(\phi)$.  The generation templates which together
parameterise all components of the $\phi$ and $\gamma$ four-momentum
vectors and the $\phi$ hadronic isolation variable are described in
table\,\ref{table:templates} and represented in
figure\,\ref{fig:sampling}.

\begin{table}[t]
  \centering
\begin{tabular}{|c|c|c|c|c|}
  \hline
    Template Name & Dimensionality & Variable 1 & Variable 2 & Variable 3 \\
    \hline
   A & 2D & $\pt(\phi)$ & $\pt(\gamma)$ & -\\
   B & 3D & $\Delta\Phi(\phi,\gamma)$ & $\pt(\gamma)$ & $\pt(\phi)$ \\
   C & 2D & $\Delta\eta(\phi,\gamma)$ & $\Delta\Phi(\phi,\gamma)$ & - \\
   D & 2D & $I(\phi)$ & $\pt(\gamma)$ & - \\
   E & 1D & $\eta(\gamma)$ &  -  & -\\
   F & 1D & $\Phi(\gamma)$ & - & - \\
   G & 1D & $m(\phi)$ & - & - \\
     \hline
\end{tabular}
\caption{The definition of the set of generation templates from which the components of the pseudo-candidates are sequentially sampled.\label{table:templates}}
\end{table}

The sampling procedure for a single pseudo-candidate proceeds as follows:

\begin{enumerate}
  \item Correlated values for $\pt(\phi)$ and $\pt(\gamma)$ are sampled from generation template A.
  \item Based on the values of $\pt(\phi)$ and $\pt(\gamma)$ sampled in step 1, template B is projected along the $\Delta\Phi(\phi,\gamma)$ dimension and a value for $\Delta\Phi(\phi,\gamma)$ is sampled.
  \item Based on the value of $\Delta\Phi(\phi,\gamma)$ sampled in step 2, template C is projected along the $\Delta\eta(\phi,\gamma)$ dimension and a value for $\Delta\eta(\phi,\gamma)$ is sampled.
  \item Based on the value of $\pt(\gamma)$ sampled in step 1, template D is projected along the $I(\phi)$ dimension and a value is sampled.
  \item Values for $\eta(\gamma)$ and $\phi(\gamma)$ are sampled from generation templates E and F, respectively. At this stage, the photon four-momentum is fully defined, with $m(\gamma)=0$ imposed.
  \item A value for $m(\phi)$ is sampled from generation template G. At this stage, the $\phi$ four-momentum is fully defined.
\end{enumerate}

The steps above are repeated to generate an ensemble of
pseudo-candidates whose characteristics resemble those of the GR data
sample.  The selection requirements of the SR and two VR are imposed
on the ensemble and the pseudo-candidates which are retained are used
to construct distributions of composite variables built from the
$\phi$ and $\gamma$ four-momentum, of which $m(\phi,\gamma)$ is of
primary interest.

\subsection{Validation}
\label{sec:validation}
\begin{figure}[!h]
\centering
  \subfigure[]{\includegraphics[width=0.4\textwidth]{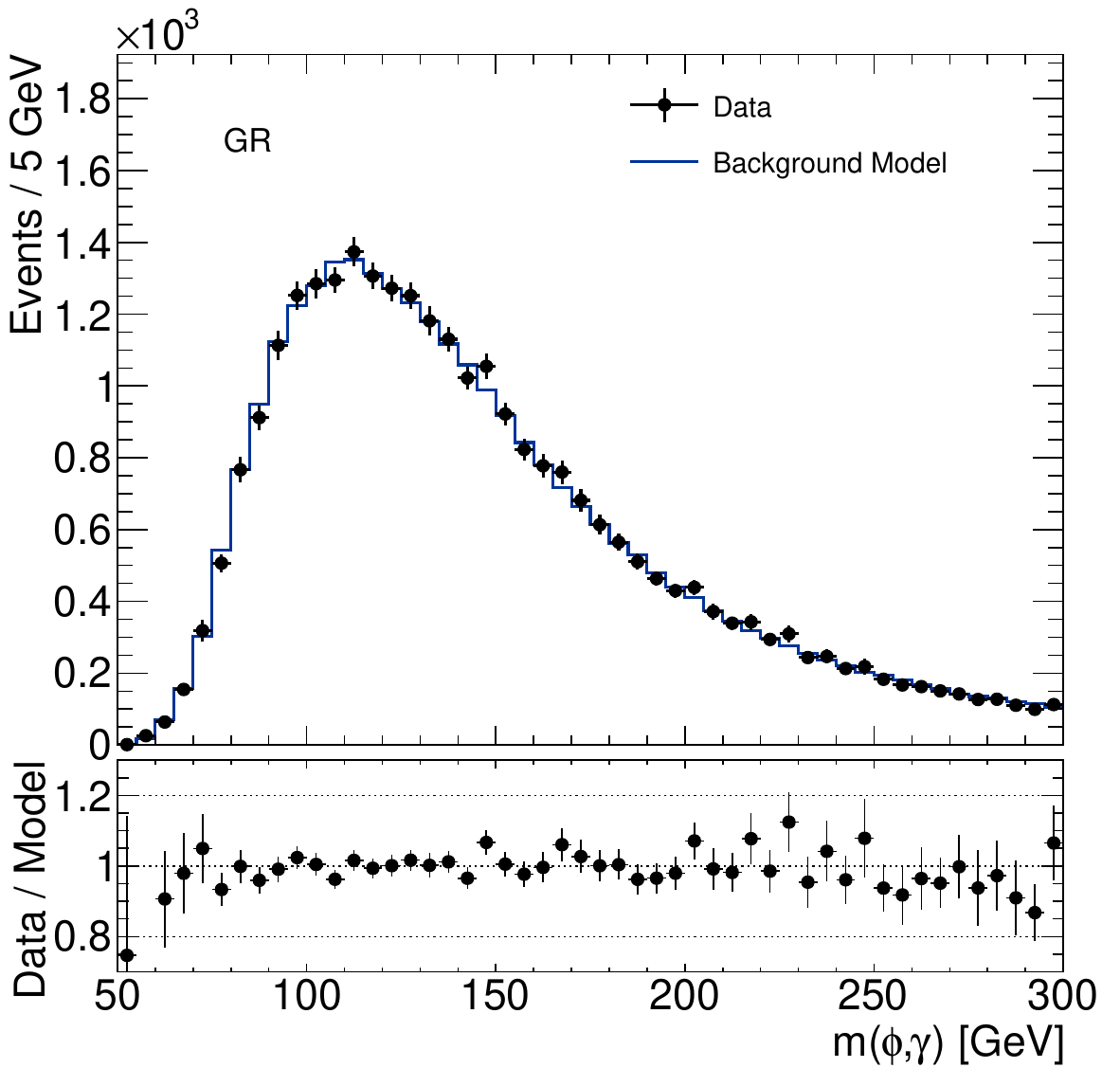}}
  \subfigure[]{\includegraphics[width=0.4\textwidth]{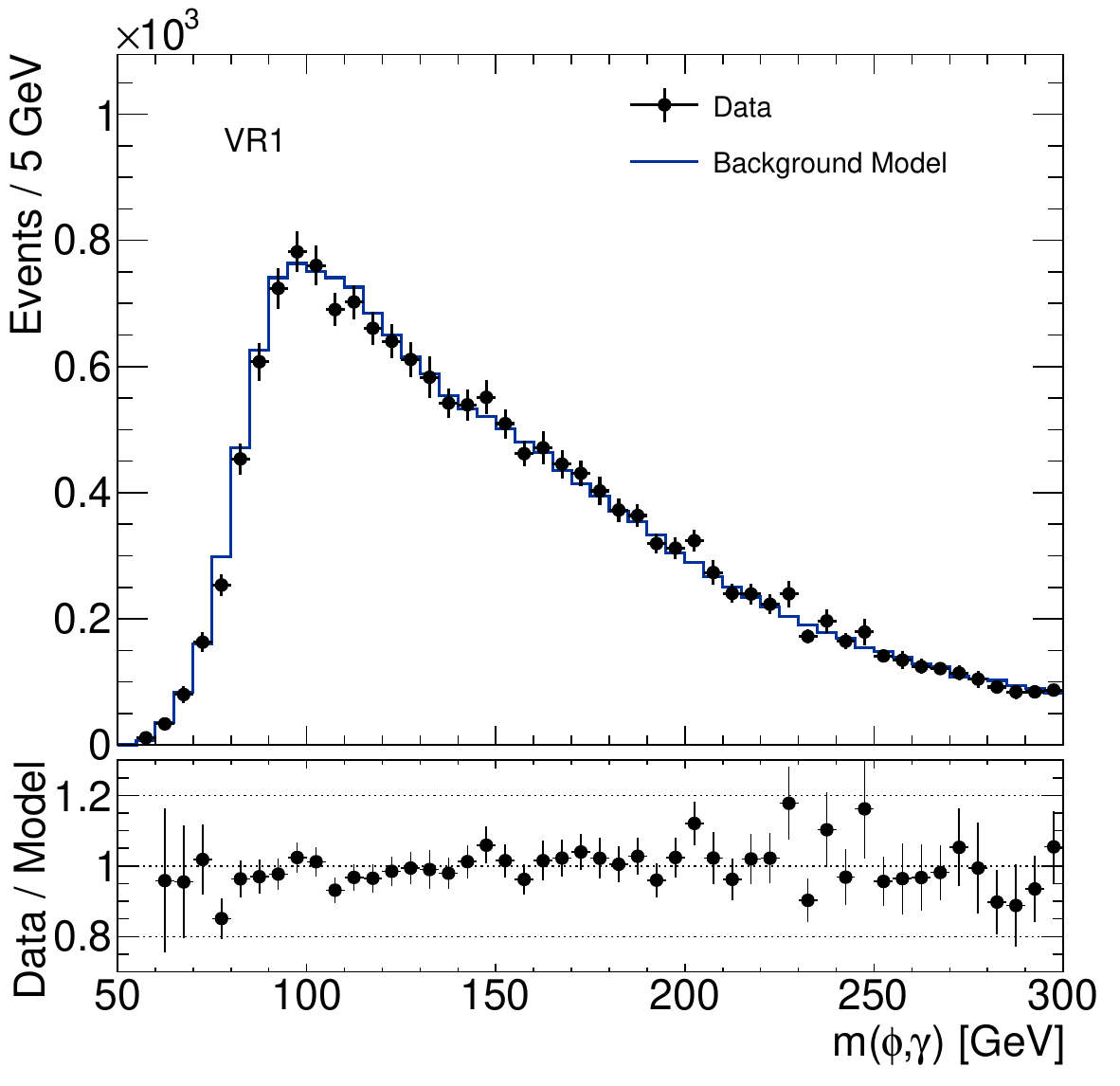}}\\
  \subfigure[]{\includegraphics[width=0.4\textwidth]{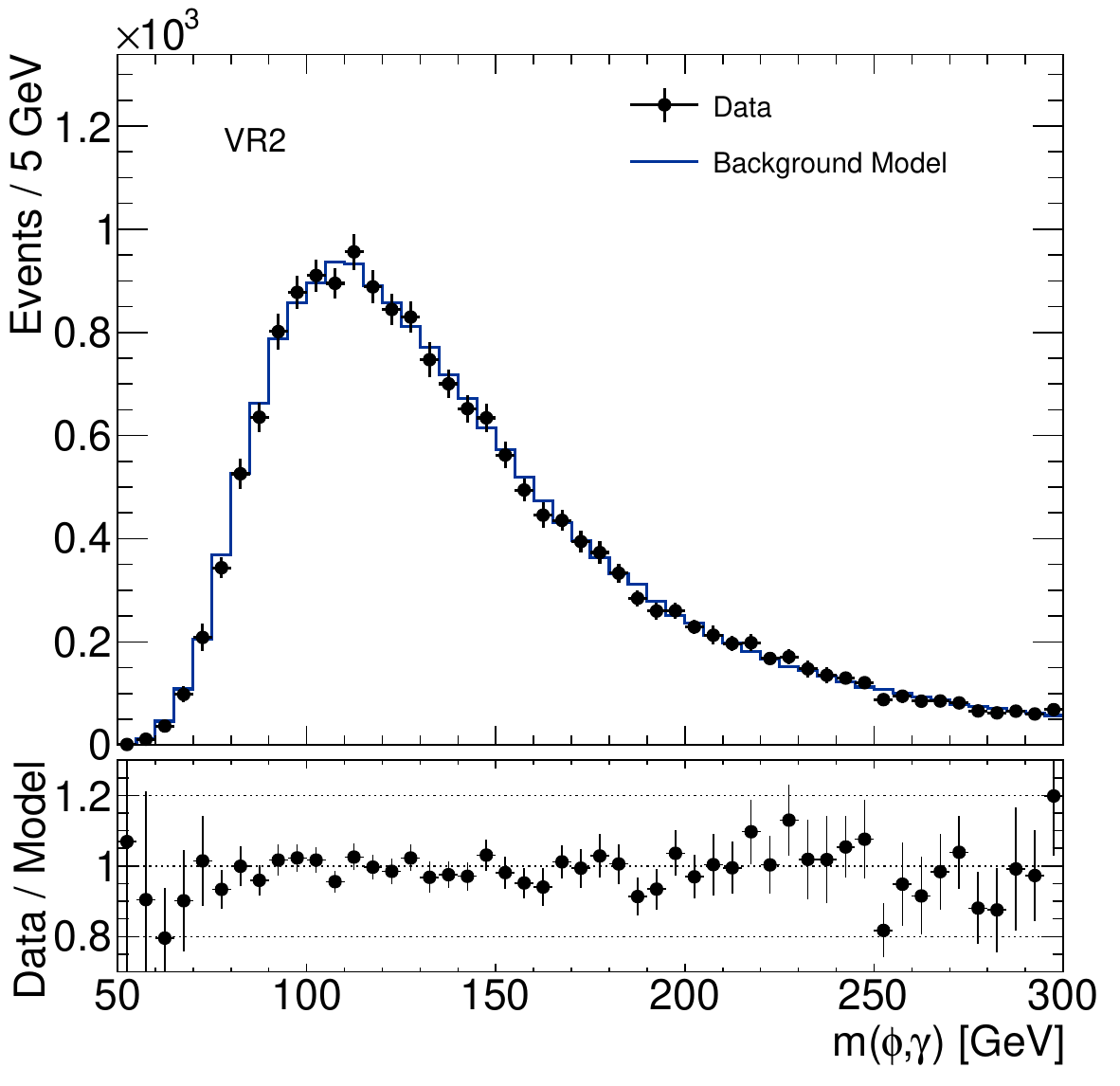}}
  \subfigure[]{\includegraphics[width=0.4\textwidth]{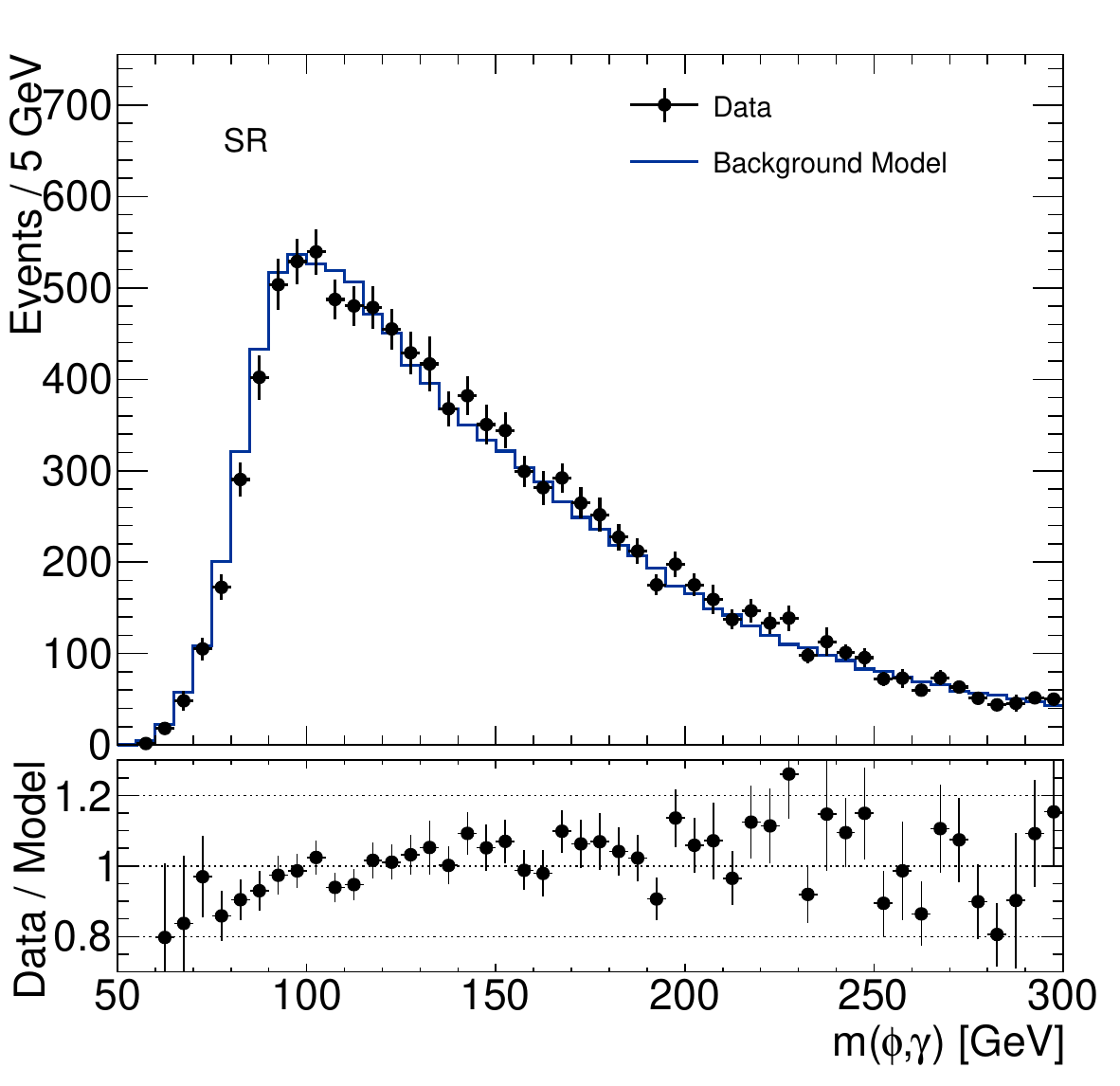}}
\caption{$m(\phi,\gamma)$ distributions in the GR, VR1, VR2 and SR for the simulated data and background model.\label{fig:bkg_Higgs_mass}}
\end{figure}

\begin{figure}[htbp!]
  \begin{center}
    \subfigure[]{\includegraphics[width=0.32\textwidth]{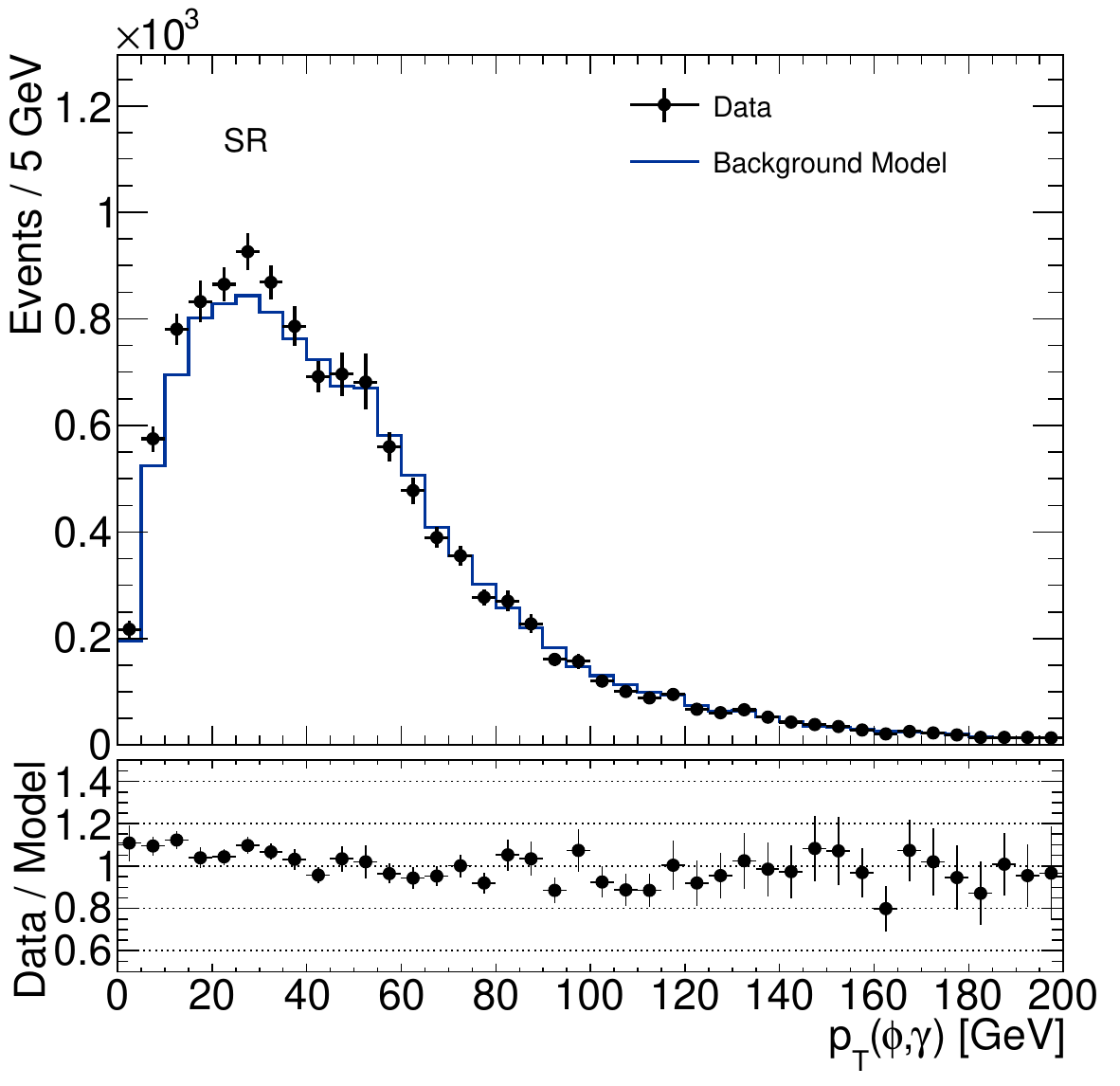}}
    \subfigure[]{\includegraphics[width=0.32\textwidth]{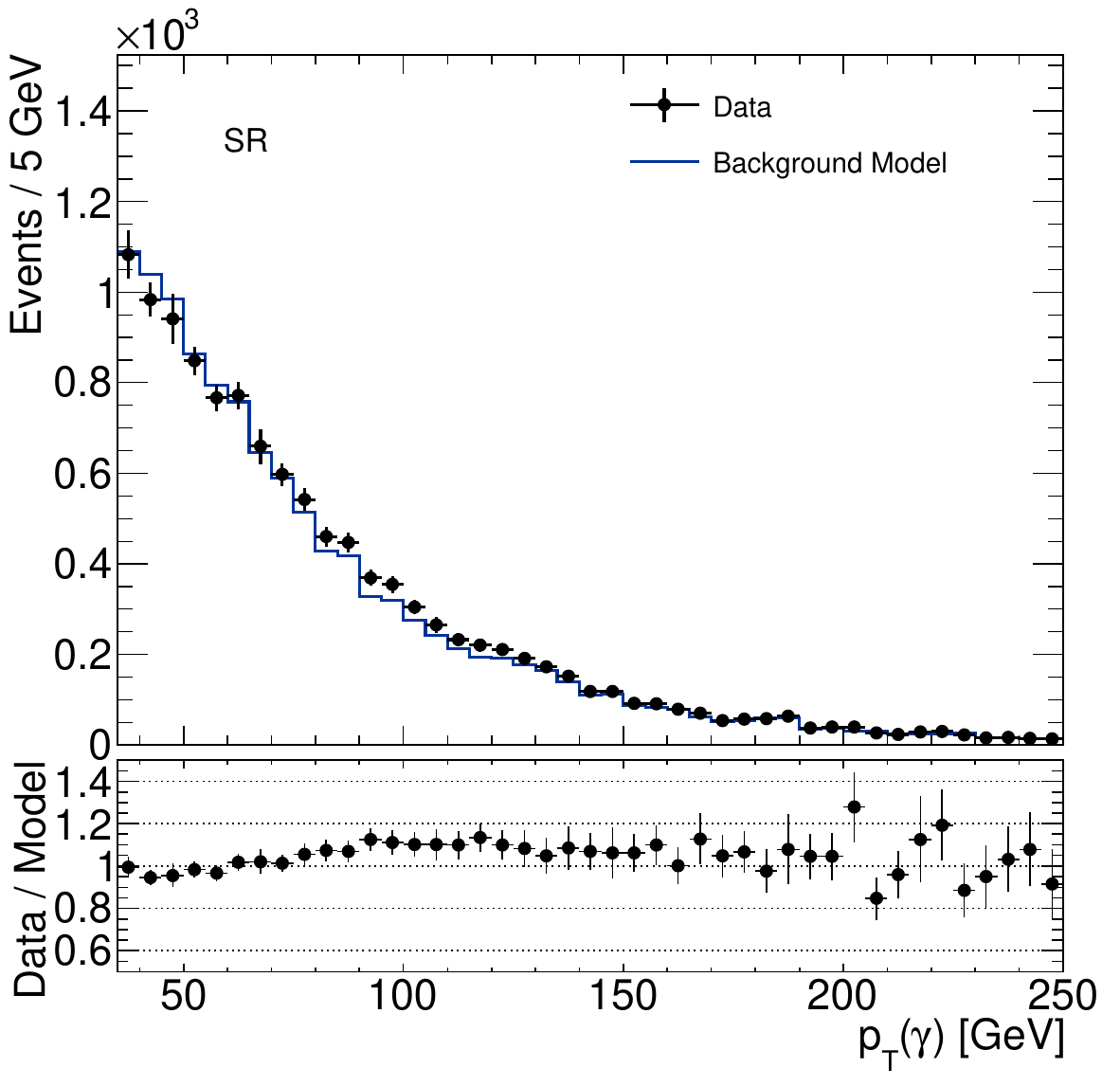}}
    \subfigure[]{\includegraphics[width=0.32\textwidth]{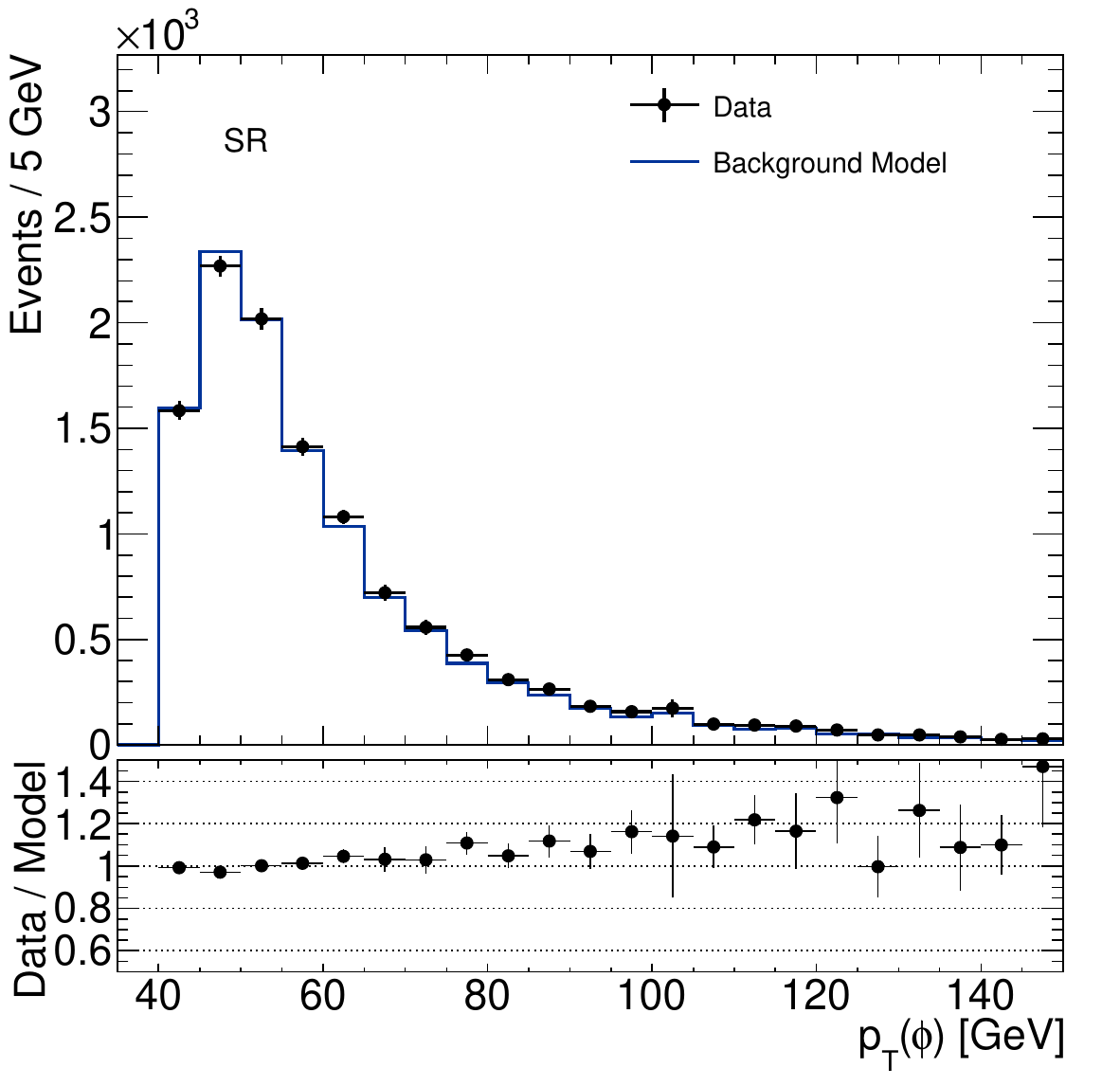}}\\
    \subfigure[]{\includegraphics[width=0.32\textwidth]{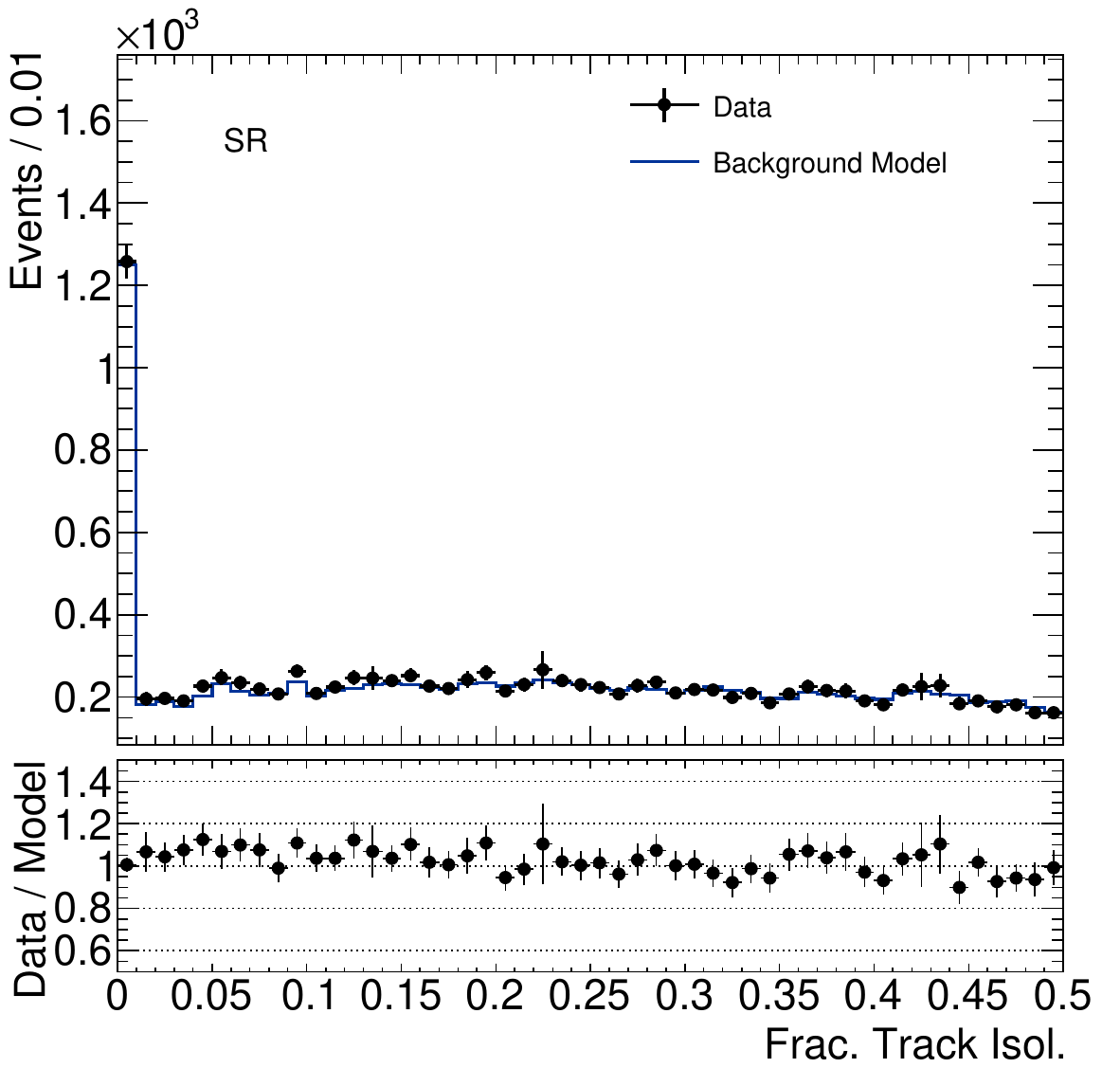}}
    \subfigure[]{\includegraphics[width=0.32\textwidth]{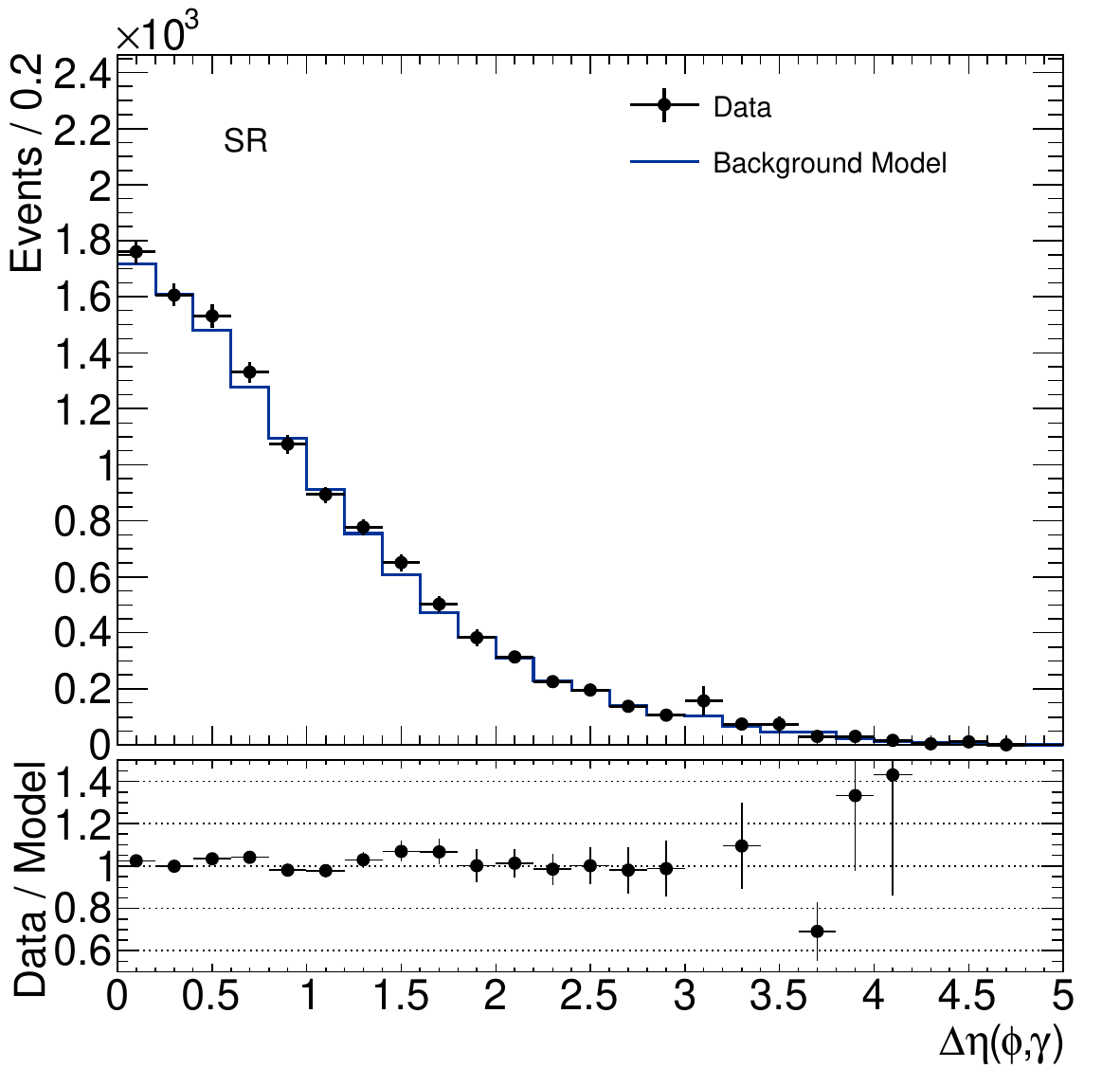}}
    \subfigure[]{\includegraphics[width=0.32\textwidth]{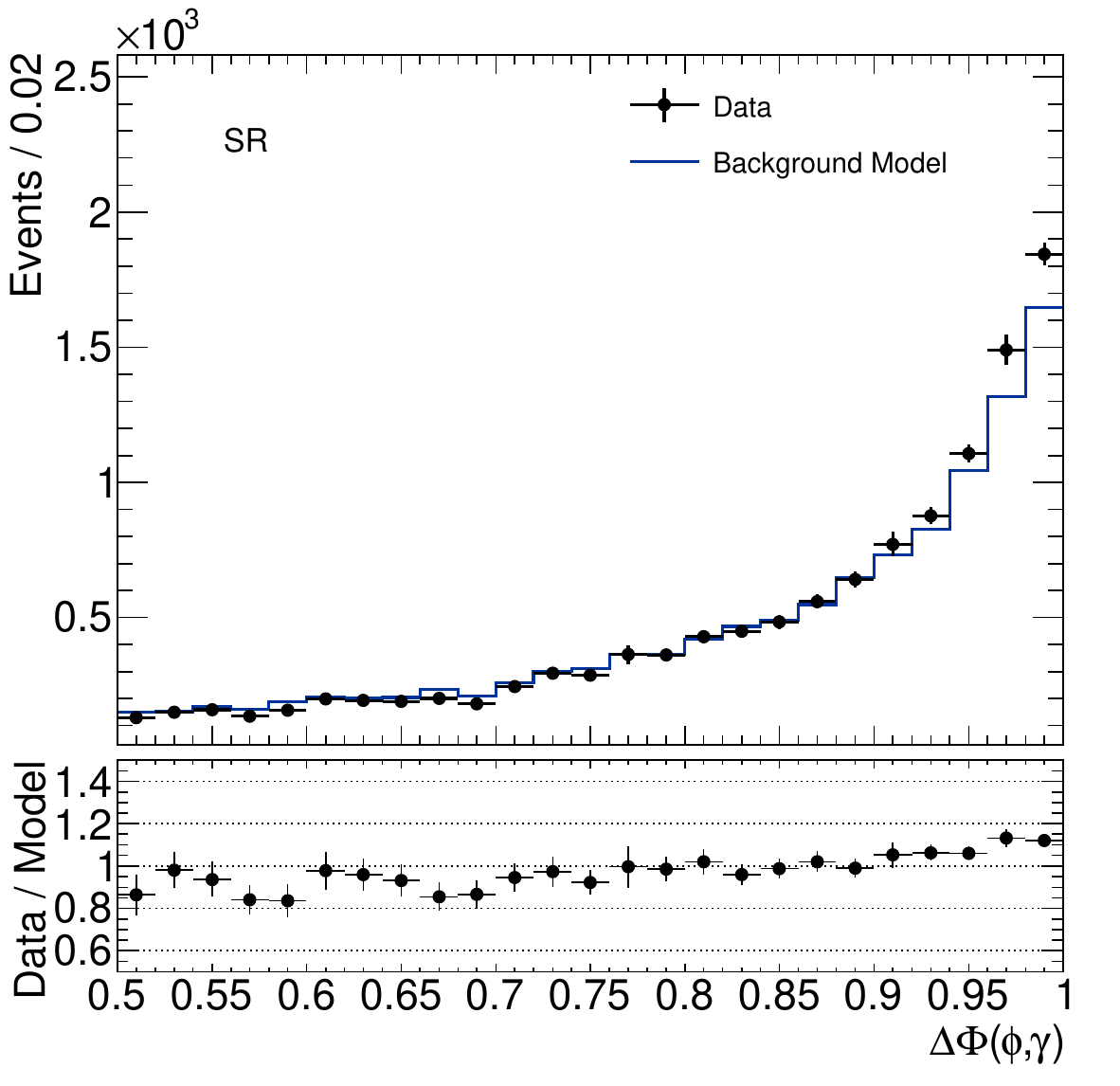}}\\
  \caption{Distributions of kinematic and isolation variables in the SR, for the simulated data and background model.\label{fig:bkg_Higgs_pt}}
  \end{center}
  \end{figure}

\begin{figure}[!htbp]
  \centering
  \subfigure[]{\includegraphics[width=0.4\textwidth]{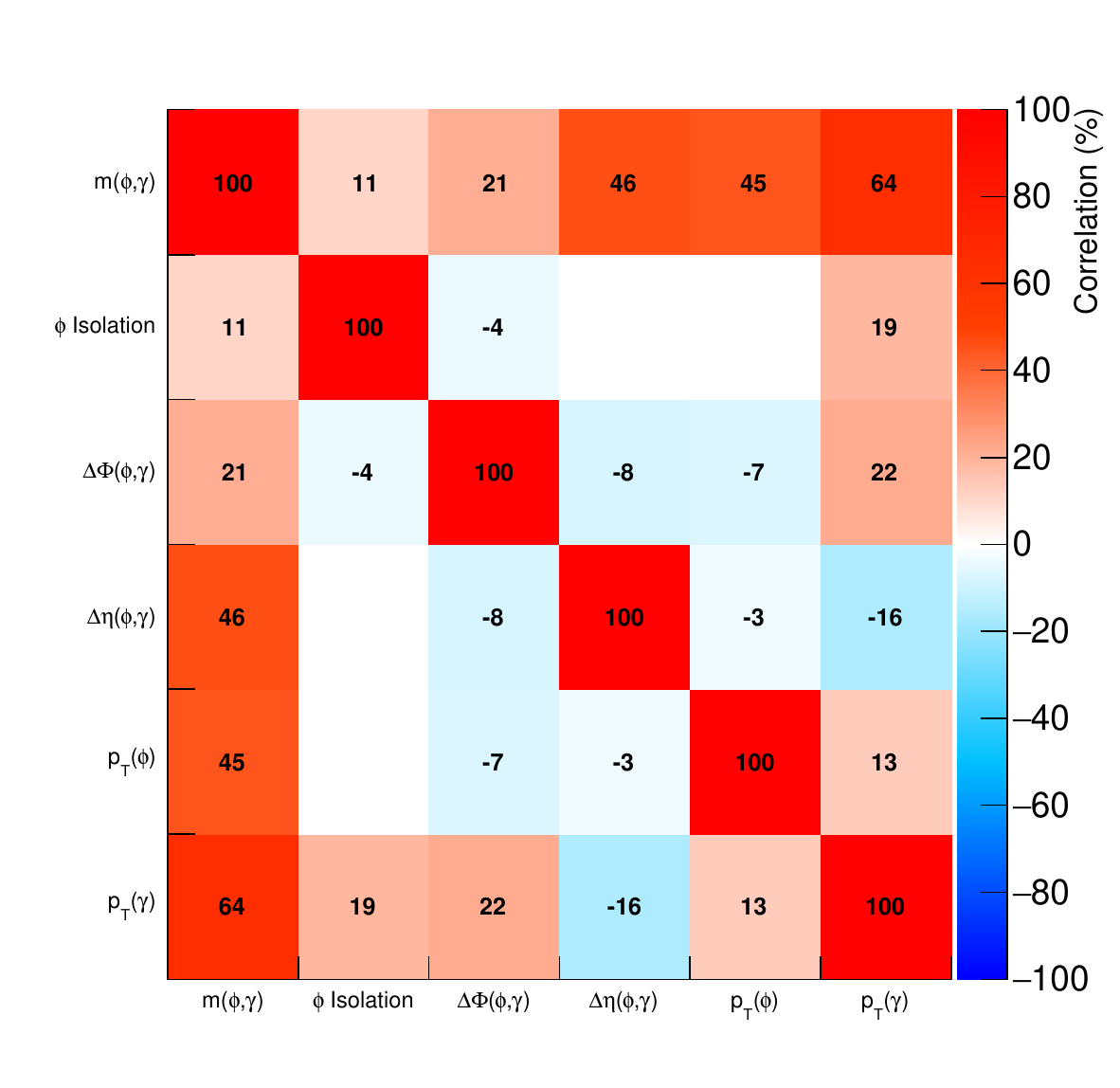}}
  \subfigure[]{\includegraphics[width=0.4\textwidth]{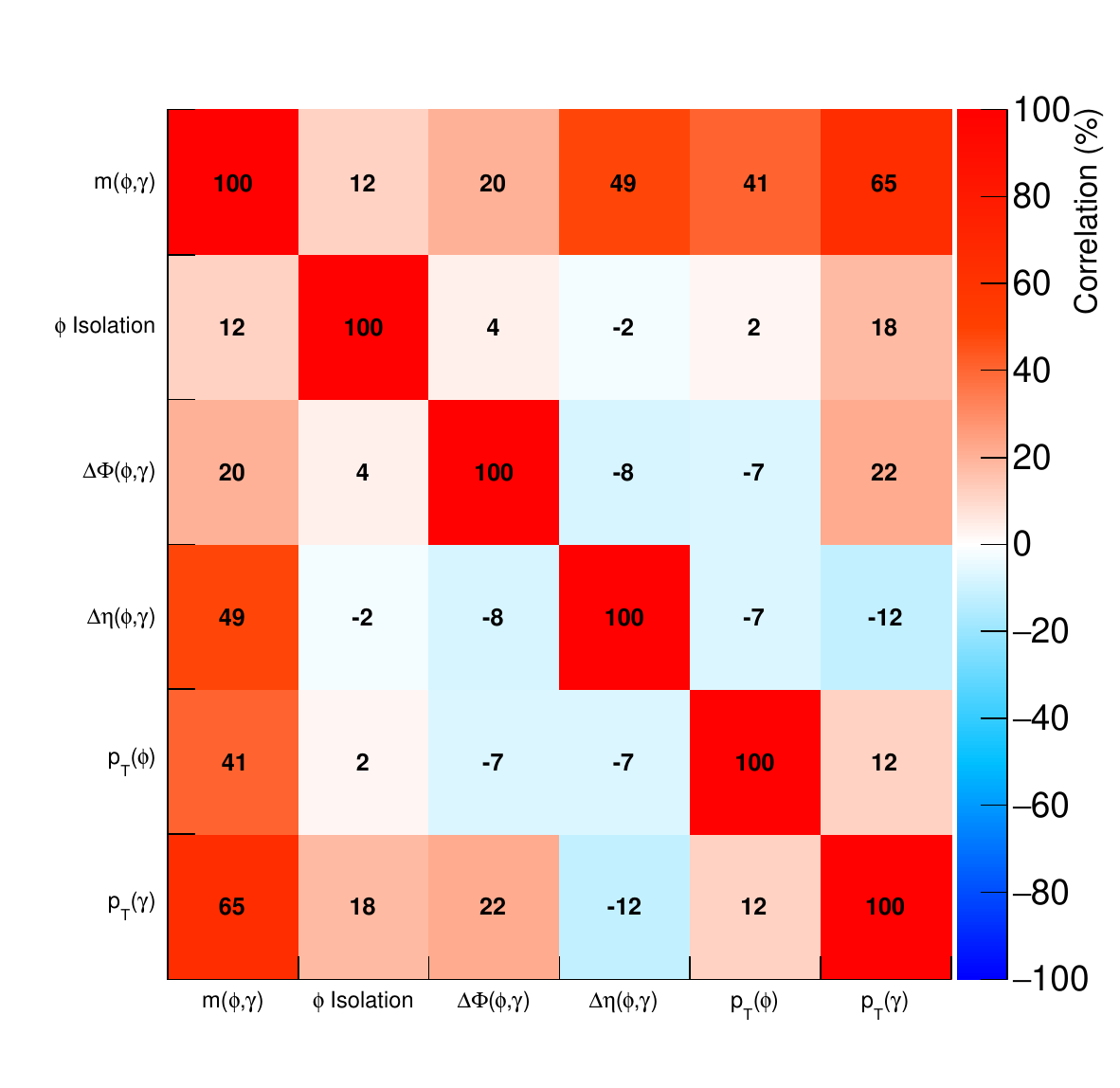}}
  \caption{Linear correlation coefficients for pairs of variables used in the background modelling procedure, shown for the simulated data events (left) and generated pseudo-candidates (right) passing the GR selections.}
  \label{fig:BkgdCorr}
\end{figure}

Both the distributions and correlations of important primary and
composite variables are validated by comparing the ensemble of
pseudo-candidates to the data sample.  Figure~\ref{fig:bkg_Higgs_mass}
shows the $m(\phi,\gamma)$ distribution of the data sample and sample
of pseudo-candidates for the GR, VR1, VR2 and SR selection criteria,
respectively. Accounting for event weighting, the GR contains 30175
events, while the SR contains 11885 events. The modelling of the
kinematic and isolation variables in the SR, for the data sample and
the sample of pseudo-candidates in the SR is presented in
figure~\ref{fig:bkg_Higgs_pt}. The accurate reproduction of composite
variables relies on the correlations among variables to be adequately
described in the ensemble of pseudo-candidates, which in turn is
determined by the sampling procedure and parameterisation of
generation templates. Figures~\ref{fig:bkg_Higgs_mass}
and\,\ref{fig:bkg_Higgs_pt} exhibit very good agreement between the
sample of pseudo-candidates and data events for both the
$m(\phi,\gamma)$ and $\pt(\phi,\gamma)$ distributions, indicating that
the important correlations in the data sample are well modelled by the
sample of pseudo-candidates.  Linear correlation coefficients,
evaluated between each of the important variables, are shown in
figure\,\ref{fig:BkgdCorr} for both the data sample and ensemble of
pseudo-candidates. Both the hierarchy and general magnitude of the
correlation coefficients in the data are well reproduced by the model.

\subsection{Signal injection tests}
\label{sec:npdd_signal}
\begin{figure}[!htbp]
  \centering
  \subfigure[\label{fig:Injection5sigma_a}]{\includegraphics[width=0.4\textwidth]{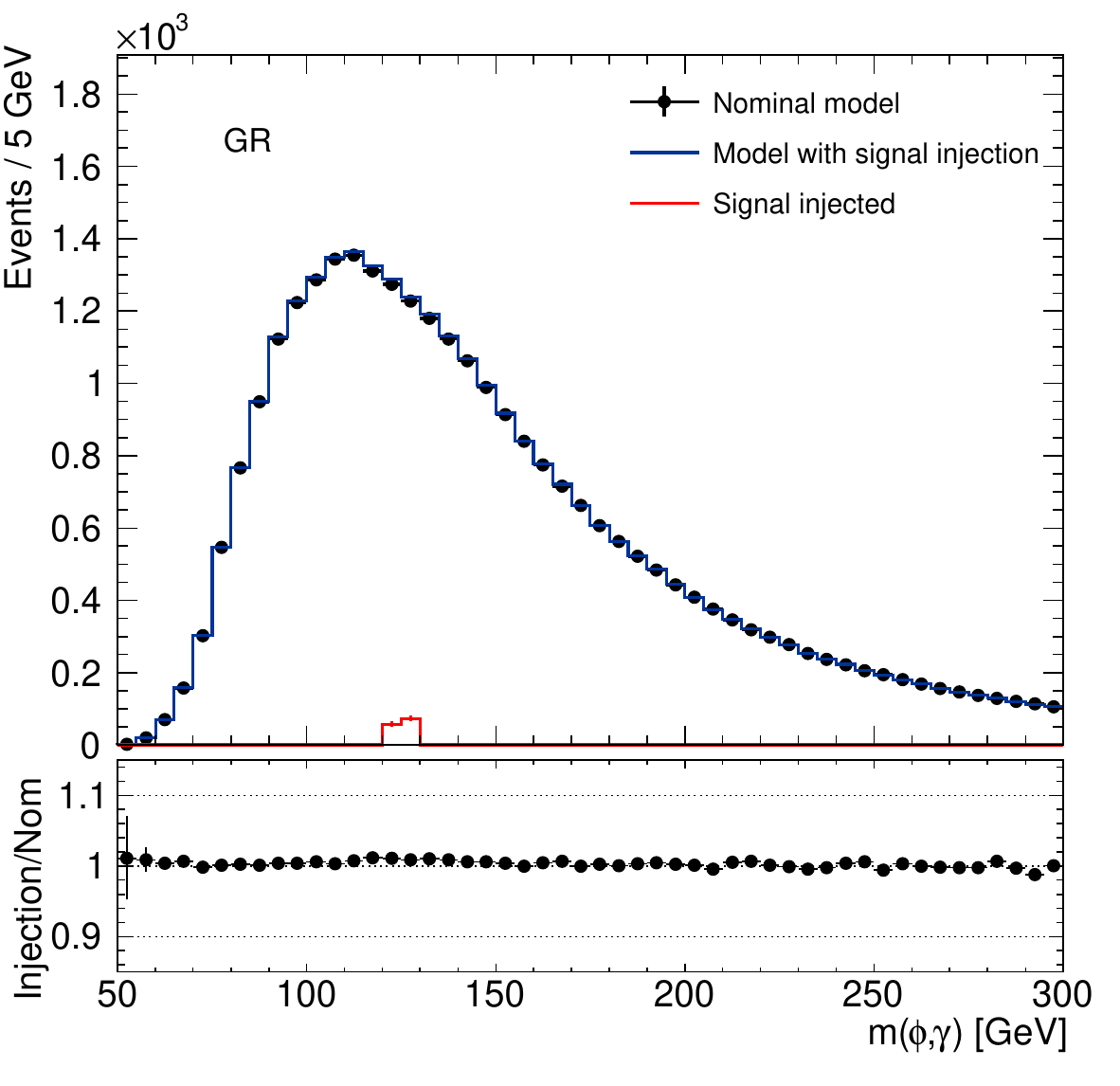}}
  \subfigure[\label{fig:Injection5sigma_b}]{\includegraphics[width=0.4\textwidth]{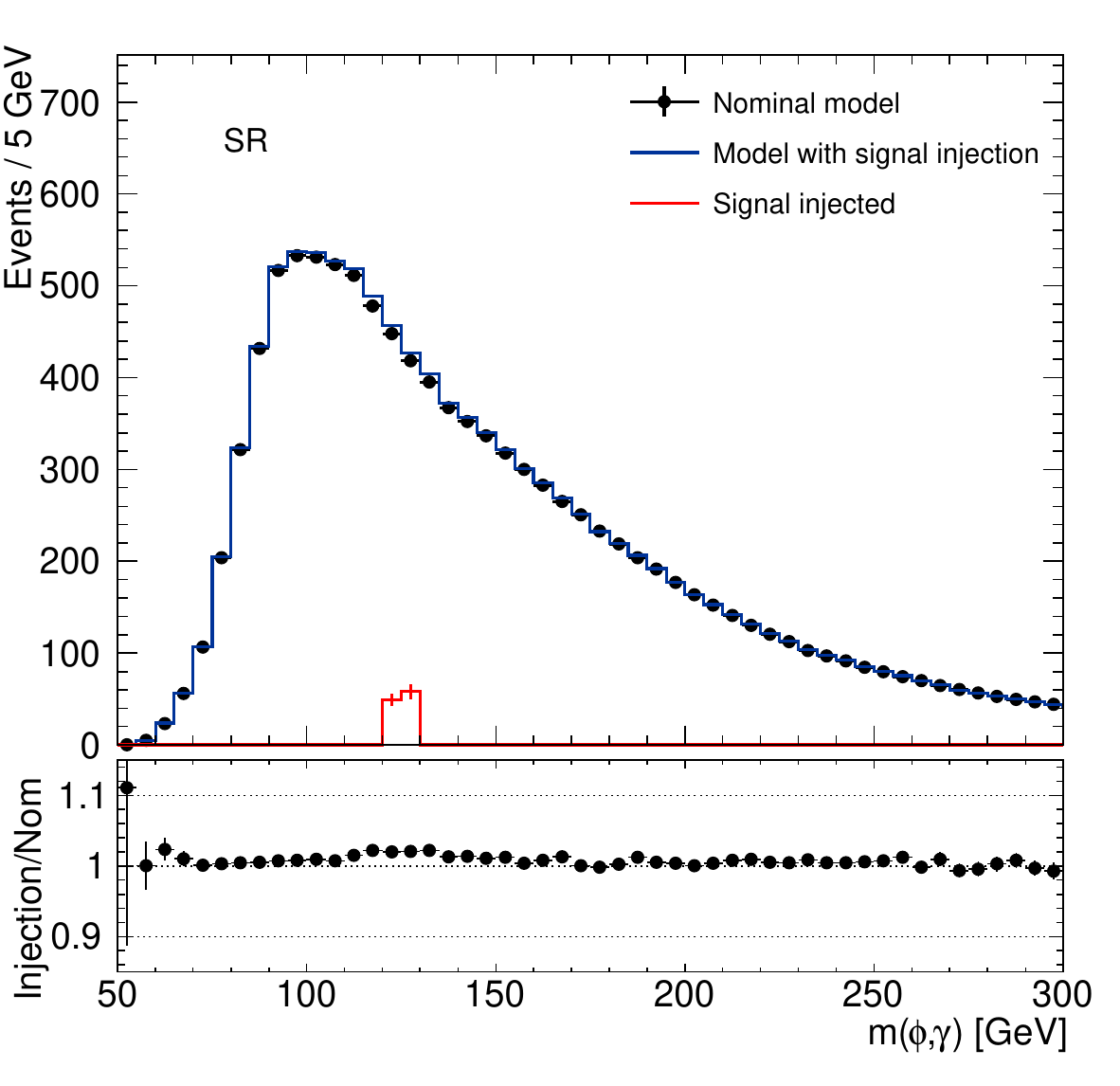}}
  \caption{Comparison of the $m(\phi,\gamma)$ distributions predicted by the model for \subref{fig:Injection5sigma_a} the GR and \subref{fig:Injection5sigma_b} the SR, derived with and without an injection of approximately 130 $H\to\phi\,\gamma$ signal events in the GR dataset.\label{fig:Injection5sigma}}
\end{figure}

While the GR dataset is defined such that it is dominated by
background events, a small contribution from signal events, relative
to the background, may be present. In the limit that the signal to
background ratio in the GR becomes large, one can intuitively expect
that the reliability of the method will begin to degrade. In order to
quantify what level of signal contribution must be present in the GR
in order to induce a noticeable impact on the reliability of the
model, signal injection tests are performed.  Specifically, it is
estimated that a signal contribution in the GR of approximately 130
events, would correspond to a statistical significance of 5.5 standard
deviations in the SR.  Consequently, 130 signal events were injected
into the GR dataset and the background model was re-derived.
Figure~\ref{fig:Injection5sigma} shows a comparison of the model
constructed with and without this signal injection.  The total
background prediction between $122.5<m(\phi,\gamma)<127.5\,\GeV$ in
the SR increases by approximately 2\%.  
The outcome of this signal injection test demonstrates that the method
is robust against the presence of even a large, in terms of
corresponding statistical significance in the SR, signal contributions
to the GR dataset. Furthermore, the effect on the background is found
to scale linearly with the number of injected signal events.

\subsection{Systematic uncertainties}
\label{sec:NPDD_systematics}
While the tests described in section~\ref{sec:validation} demonstrate
that the model can provide an accurate description of the background
at the level of precision associated with the statistical uncertainty
of the validation regions, potential mismodelling beyond this level
cannot be directly excluded. It is therefore important that systematic
uncertainties affecting the shape of the predicted background
distributions are estimated.

The strategy for incorporating systematic uncertainties in the model
is motivated by the context within which the model is applied to
perform the statistical analysis, namely a likelihood fit. The
strategy focuses on estimating a set of complementary background shape
variations which are implemented as shape variations of the nominal
background probability density function. The derived shape variations
are selected to capture different modes of potential deformations of
the background shape. The exact size of the variation is of less
importance, given that the corresponding nuisance parameters are
constrained directly by the data in the likelihood fit. Moreover, the
analyser may decide to leave such nuisance parameters completely free,
or to add Gaussian constraint terms in the likelihood.  In the latter
case, care must be taken to ensure that the assigned $\pm 1\sigma$
shape variations are large with respect to potential discrepancies
between the shape of the predicted distributions and those in the
data.

Pairs of approximately anti-symmetric shape variations are built by
performing the sampling procedure after having applied a
transformation to one of the generation templates. The transformations
considered here include:

\begin{itemize}
\item A translation of the photon transverse momentum distribution.
\item A multiplicative transformation of the $\Delta\Phi(\phi,\gamma)/\pi$ distribution by a function of the form $1+C\times\Delta\Phi(\phi,\gamma)$, where a pair of values (positive and negative) for the coefficient $C$ are chosen.
\end{itemize}

\begin{figure}[!htbp]
  \centering
  \subfigure[]{\includegraphics[width=0.4\textwidth]{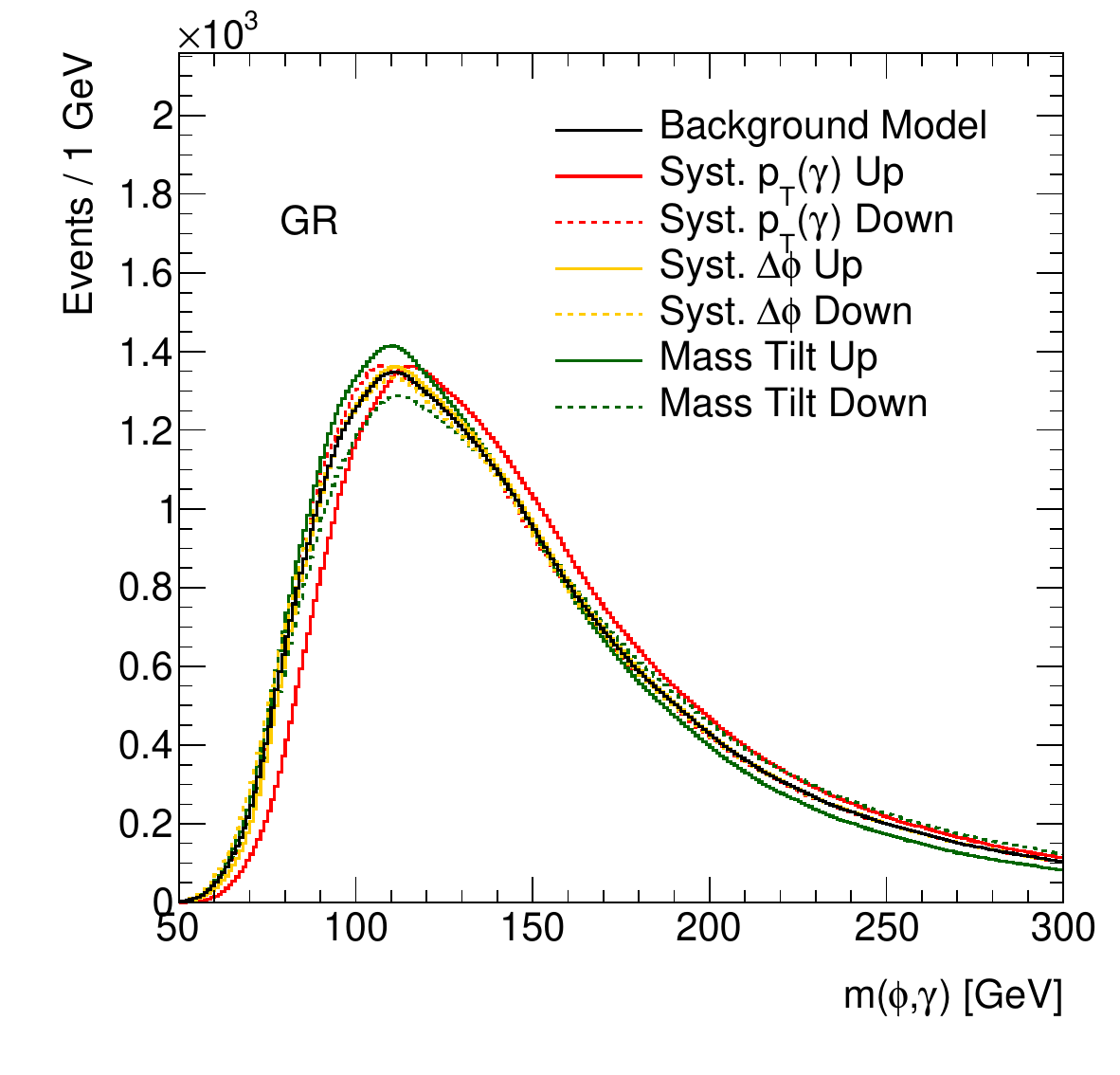}}
  \subfigure[]{\includegraphics[width=0.4\textwidth]{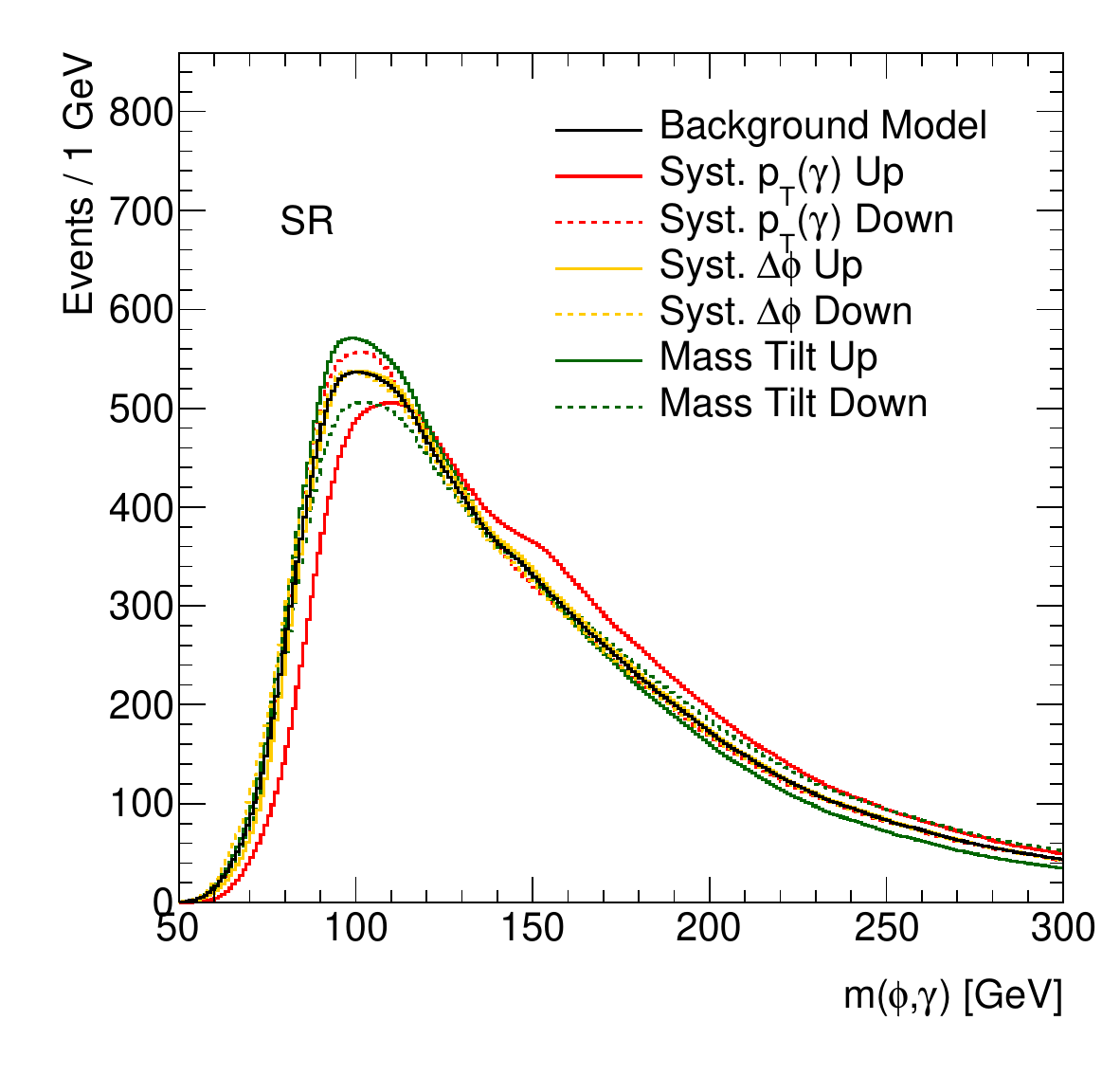}}
  \caption{Comparison of the $m(\phi,\gamma)$ distributions for the GR (left) and SR (right) associated with the three pairs of systematic shape variations. }
  \label{fig:shapesyst}
\end{figure}

\noindent Furthermore, additional alternative background models may be
derived by direct transformations of the resulting distribution of
interest. For example, an additional pair of alternative background
models is derived by a multiplicative transformation of SR
$m(\phi,\gamma)$ distribution by a linear function of
$m(\phi,\gamma)$.  These three pairs of shape variations largely span
the range of possible large scale uncorrelated shape deformations of
the $m(\phi,\gamma)$ distribution, as shown in
figure~\ref{fig:shapesyst}.

\subsection{Treatment of resonant background components}
In certain cases, in addition to the dominant backgrounds which do not
exhibit resonant structures in the invariant mass distribution,
non-negligible resonant contributions may also be present. One such
example is the $Z\to\mu^{+}\mu^{-}\gamma$ process, which represents an
important resonant background in the case of searches for radiative
Higgs boson decays to the $\Upsilon(\to\mu^{+}\mu^{-})$ bottomonium
states~\cite{Aad:2015sda,Aaboud:2018txb}.  Often such processes can be
described with sufficient accuracy by MC simulations and the use of
such simulations to describe these subsets of the overall background
is preferable. In this situation, the procedure used to build the
kernel is modified, to ensure that such resonant contributions are not
included in the model for the inclusive non-resonant
background. During the construction of the generation templates, data
events in the vicinity of the resonance are randomly discarded, with a
probability which describes the likelihood that a data event with a
given invariant mass was produced by the resonant process.  This
probability may be determined, as a function of invariant mass, by
subtracting the distribution of the MC prediction for the resonant
process from the data distribution. This procedure can mitigate the
impact of resonant background contributions to the non-resonant
background model~\cite{Aad:2015sda,Aaboud:2018txb}.  In the absence of
such a procedure, the prediction of the model in the vicinity of the
resonance may be distorted in a manner similar to the signal inject
tests discussed in section~\ref{sec:npdd_signal}.

\subsection{Implementation in statistical analysis}
\label{NPDD_stat}
The performance of the method in practical terms is demonstrated by
implementing the background model within a statistical analysis
procedure. A binned maximum likelihood fit is performed to the
$m(\phi,\gamma)$ distribution of the simulated $\gamma+\text{jet}$
events used to build the background model alone.  The $m(\phi,\gamma)$
distribution of the signal is modelled by a double Gaussian
distribution with a single mean (common to both Gaussian components),
two width parameters and a parameter describing the relative
normalisation of the two Gaussian components. The normalisation of the
signal, relative to the number of signal events predicted by the
simulation, is controlled by a single free parameter,
$\mu_{\text{signal}}$, normalised such that $\mu_{\text{signal}}=1$
corresponds to 50 signal events.  The background distribution is built
from the background model in terms of a finely binned histogram, with
a linear inter-bin interpolation applied. For reference the generated
background sample was approximately a factor 300 larger than the GR
dataset. The normalisation of the background, relative to the number
of events predicted by the model, is determined by a single free
parameter, $\mu_{\text{bkgd}}$.  Systematic uncertainties affecting
the shape of the background model, described in
section~\ref{sec:NPDD_systematics}, are implemented using a moment
morphing technique~\cite{Baak:2014fta}. Each of the three shape
variations described in section~\ref{sec:NPDD_systematics} are
implemented independently, each being controlled by an individual
nuisance parameter. The value of the nuisance parameters describing
the $\pt(\gamma)$ shift and $\Delta\Phi(\phi,\gamma)$ deformation are
constrained by Gaussian penalty terms in the likelihood, while the
$m(\phi,\gamma)$ tilt nuisance parameter is free.

\begin{figure}[!h]
  \centering
  \includegraphics[width=0.45\textwidth]{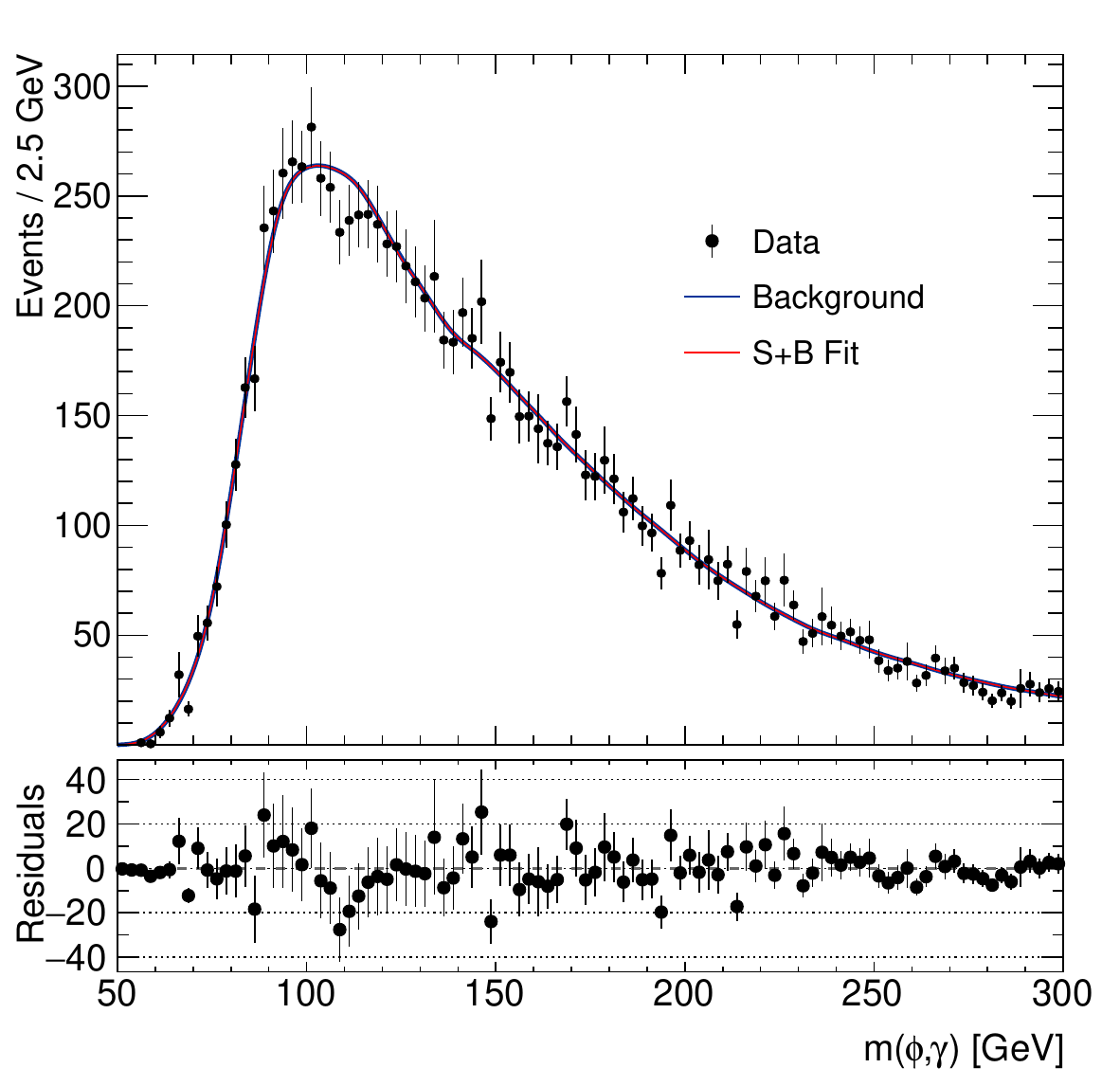}
  \caption{The $m(\phi,\gamma)$ distribution of simulated $\gamma+\text{jet}$ events overlaid with the result of a binned maximum likelihood fit using the method described in section~\ref{sec:npdd} to derive the background PDF.\label{fig:fit_npdd}}
\end{figure}

The result of the fit is presented in figure~\ref{fig:fit_npdd}, where
it is shown that the post-fit background model is in good agreement
with the $m(\phi,\gamma)$ distribution of the simulated
$\gamma+\text{jet}$ events. The fitted values of the parameters are
shown in table~\ref{fig:fit_npdd_table}.  As discussed earlier, the
post-fit uncertainty of the nuisance parameters indicates the
statistical power of the data to constrain the corresponding shape
deformations with respect to their pre-fit magnitude. Furthermore, the
non-zero post-fit values of the nuisance parameters demonstrate the
ability of the background model to absorb potential residual
mismodelling effects.

\begin{table}[h]
\centering
\begin{tabular}{|c|c|c|}
\hline
Parameter                                                            & Value & Uncertainty ($\pm1\sigma$)    \\
\hline
$\mu_{\text{signal}}$                                        & $-0.07$    &  $\pm0.54$  \\
\hline
$\mu_{\text{bkgd}}$       & $1.01$     & $\pm0.01$          \\
\hline
Shape: $\pt(\gamma)$ shift & $0.26$     & $\pm0.15$          \\
\hline
Shape: $\Delta\Phi(\phi,\gamma)$ tilt & $0.30$     & $\pm0.43$          \\
\hline
Shape: $m(\phi,\gamma)$ tilt & $0.10$     & $\pm0.24$          \\
\hline
\end{tabular}
\caption{Post-fit parameter values and their associated uncertainties.\label{fig:fit_npdd_table}}
\end{table}

The impact of the systematic uncertainties on the signal strength
measurement is quantified through fits in which the nuisance
parameters associated with the shape variations are fixed, one at a
time, to their best fit values plus or minus their correspond
uncertainty. The change in the obtained $\mu_{\text{signal}}$ in each
of these fits, expressed in terms of $\sigma_{\mu_\text{signal}}$, is
shown in table~\ref{tab:syst_effect}.

\begin{table}[!htbp]
\centering
\begin{tabular}{cc|c|}
\cline{3-3}
                                                                             &            & $\Delta\mu_{\text{signal}}/\sigma_{\mu_{\text{signal}}}$ \\ \hline
\multicolumn{1}{|c|}{\multirow{2}{*}{Shape: $\pt(\gamma)$ shift}}            & $+1\sigma$ &-0.18                   \\ \cline{2-3} 
\multicolumn{1}{|c|}{}                                                       & $-1\sigma$ &+0.17                   \\ \hline
\multicolumn{1}{|c|}{\multirow{2}{*}{Shape: $\Delta\Phi(\phi,\gamma)$ tilt}} & $+1\sigma$ &+0.07                   \\ \cline{2-3} 
\multicolumn{1}{|c|}{}                                                       & $-1\sigma$ &-0.11                   \\ \hline
\multicolumn{1}{|c|}{\multirow{2}{*}{Shape: $m(\phi,\gamma)$ tilt}}          & $+1\sigma$ &-0.22                   \\ \cline{2-3} 
\multicolumn{1}{|c|}{}                                                       & $-1\sigma$ &+0.19                   \\ \hline
\end{tabular}
\caption{The change in the fitted signal strength, relative  to its uncertainty, is shown for the case where the corresponding nuisance parameter is fixed to its best fit value plus or minus its corresponding uncertainty. \label{tab:syst_effect}}
\end{table}

\subsection{Ensemble tests using synthetic datasets}
\label{sec:NPDD_toys}

In order to more thoroughly validate the performance of the method in
a manner less sensitive to statistical fluctuations, an ensemble of
independent statistical tests identical to that described in
section~\ref{NPDD_stat} were performed. For the purposes of these
tests, a synthetic dataset sampled from analytic functions was
generated in order to simulate an appropriately large number of events
on a practical timescale. The form of the analytic functions was
chosen to mimic the relevant distributions of the MC simulated events
described in section~\ref{analysis_phiy}. Events with a complete
four-vector for the $\phi\gamma$ system are generated according to the
following procedure:

\begin{itemize}
    \item A value for $m(\phi,\gamma)$ is sampled from a Landau distribution, whose parameters are chosen to mimic the GR distribution in figure~\ref{fig:bkg_Higgs_mass}
    \item A value for $p_{\text{T}}(\phi,\gamma)$ is sampled from a Gaussian distribution, whose mean and standard deviation parameters vary as a function of $m(\phi,\gamma)$
    \item A value for $\eta(\phi,\gamma)$ is sampled from an analytical distribution $f(\eta) = G(\eta,-\mu,\sigma) + G(\eta,\mu,\sigma)$, where $G$ is the Gaussian distribution
    \item A value for $\Phi(\phi,\gamma)$ is uniformly sampled in the interval $\Phi(\phi,\gamma)\in\{-\pi,\pi\}$

\end{itemize}

Four-vectors for the $\phi$ (and subsequent $\phi\to K^{+}K^{-}$ decay
products) and $\gamma$ are generated by means of kinematic phase space
sampling with isotropic angular dependence. Finally, a $\phi$
isolation variable sampled from a linear function of
$p_{\text{T}}(\phi)$, chosen to mimic the distribution of the MC
simulated events described in section~\ref{analysis_phiy}.

An ensemble of 400 independent background-only samples was generated,
each containing around $23,000$ synthetic background events in the GR
and $16,000$ events in the SR. The number of synthetic background
events in each ensemble was chosen to approximately match the
effective statistical power of the background MC sample described in
section~\ref{analysis_phiy}.

\begin{figure}[]
\centering
  \subfigure[\label{fig:npdd_pull_a}]{\includegraphics[width=0.45\textwidth]{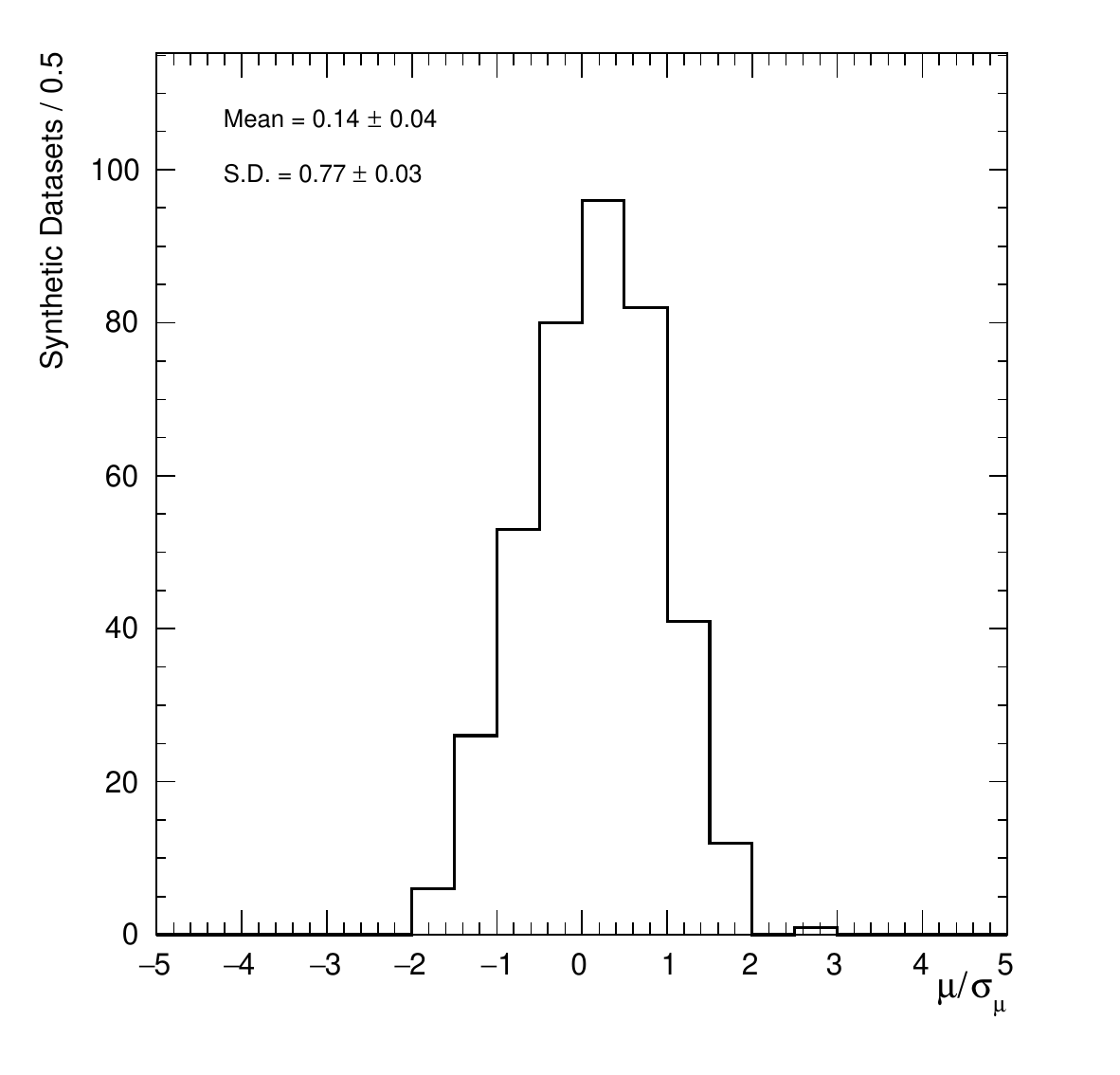}}
  \subfigure[\label{fig:npdd_pull_b}]{\includegraphics[width=0.45\textwidth]{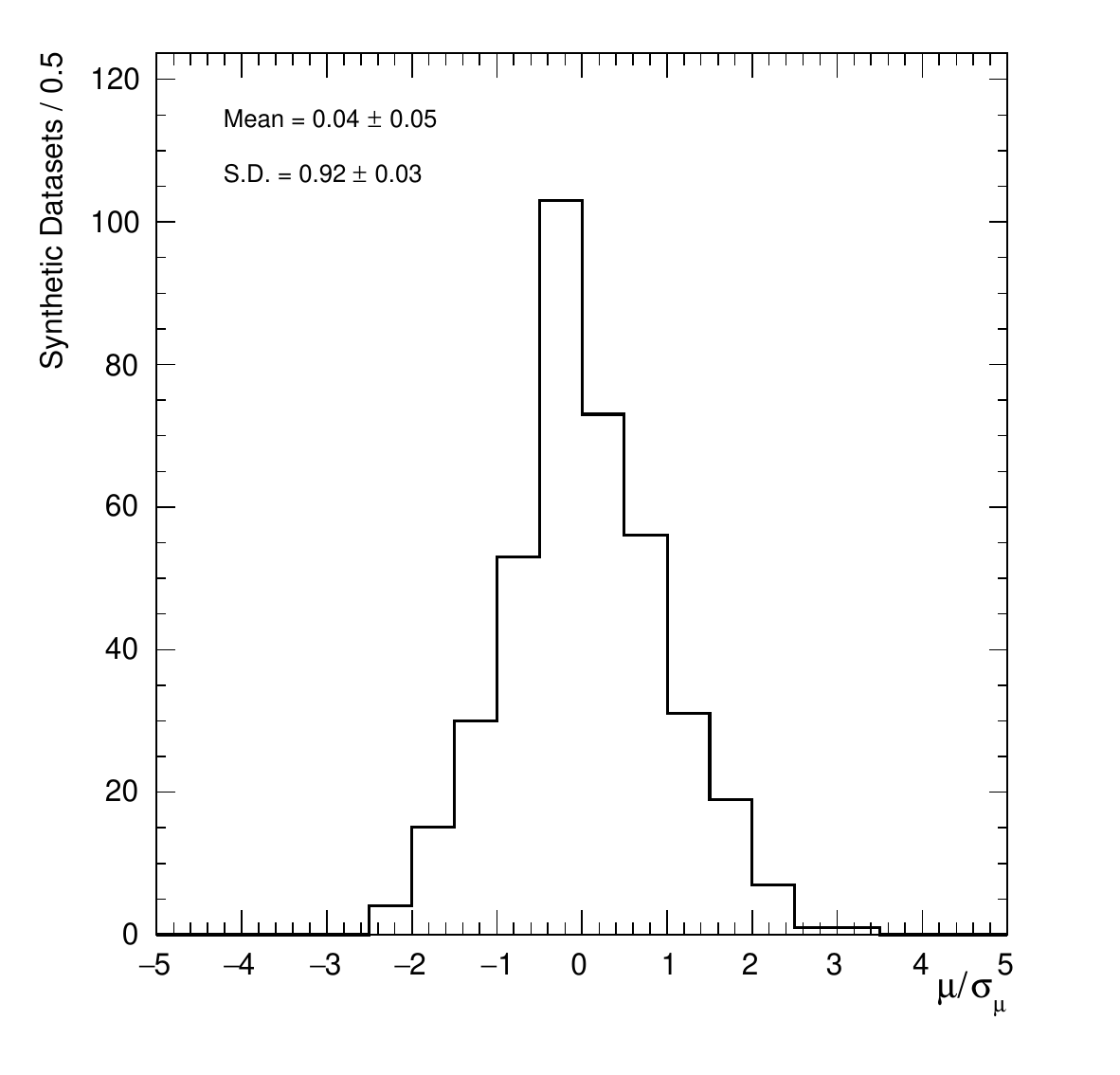}}\\
  \subfigure[\label{fig:npdd_pull_c}]{\includegraphics[width=0.45\textwidth]{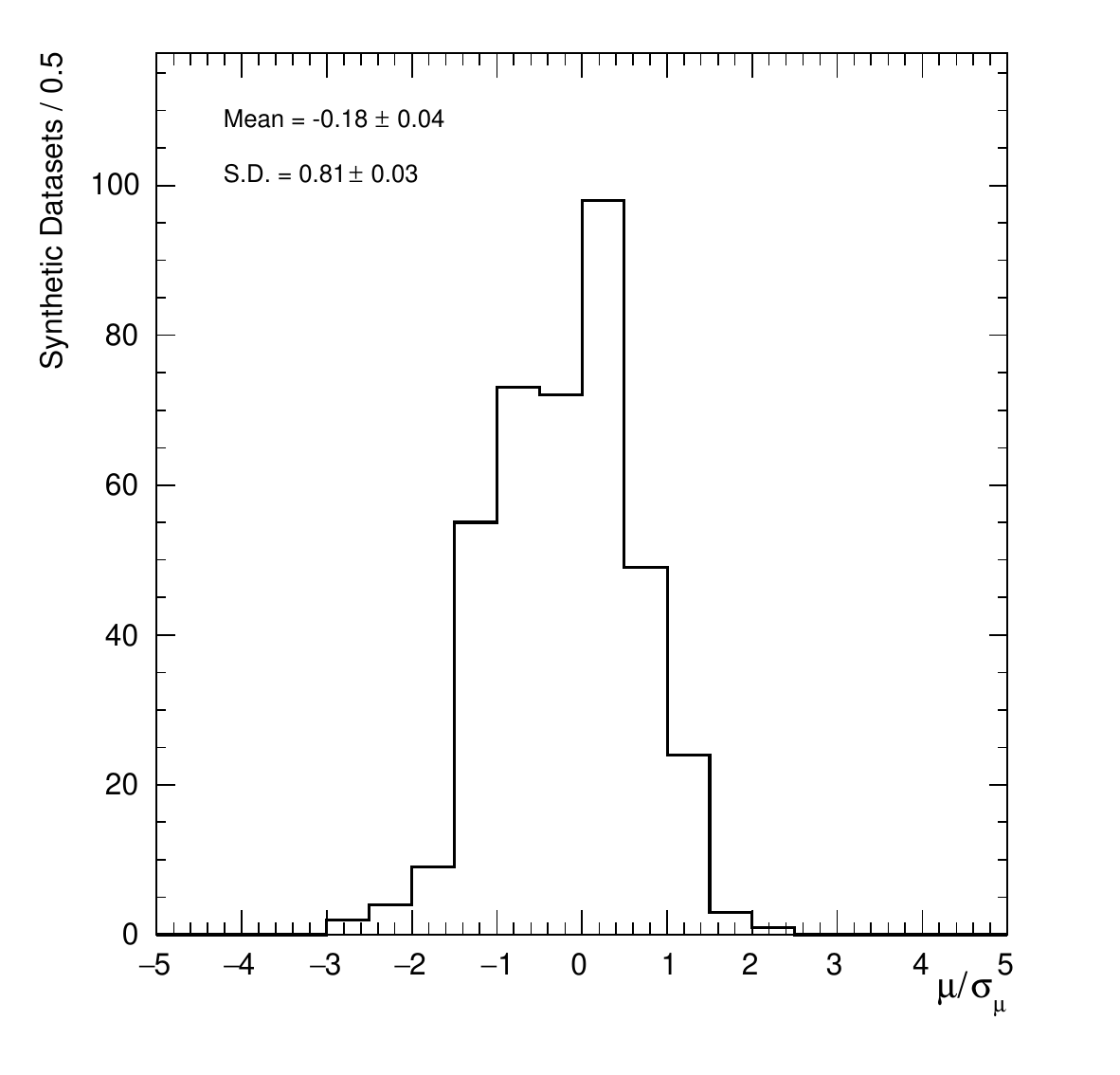}}
  \subfigure[\label{fig:npdd_pull_d}]{\includegraphics[width=0.45\textwidth]{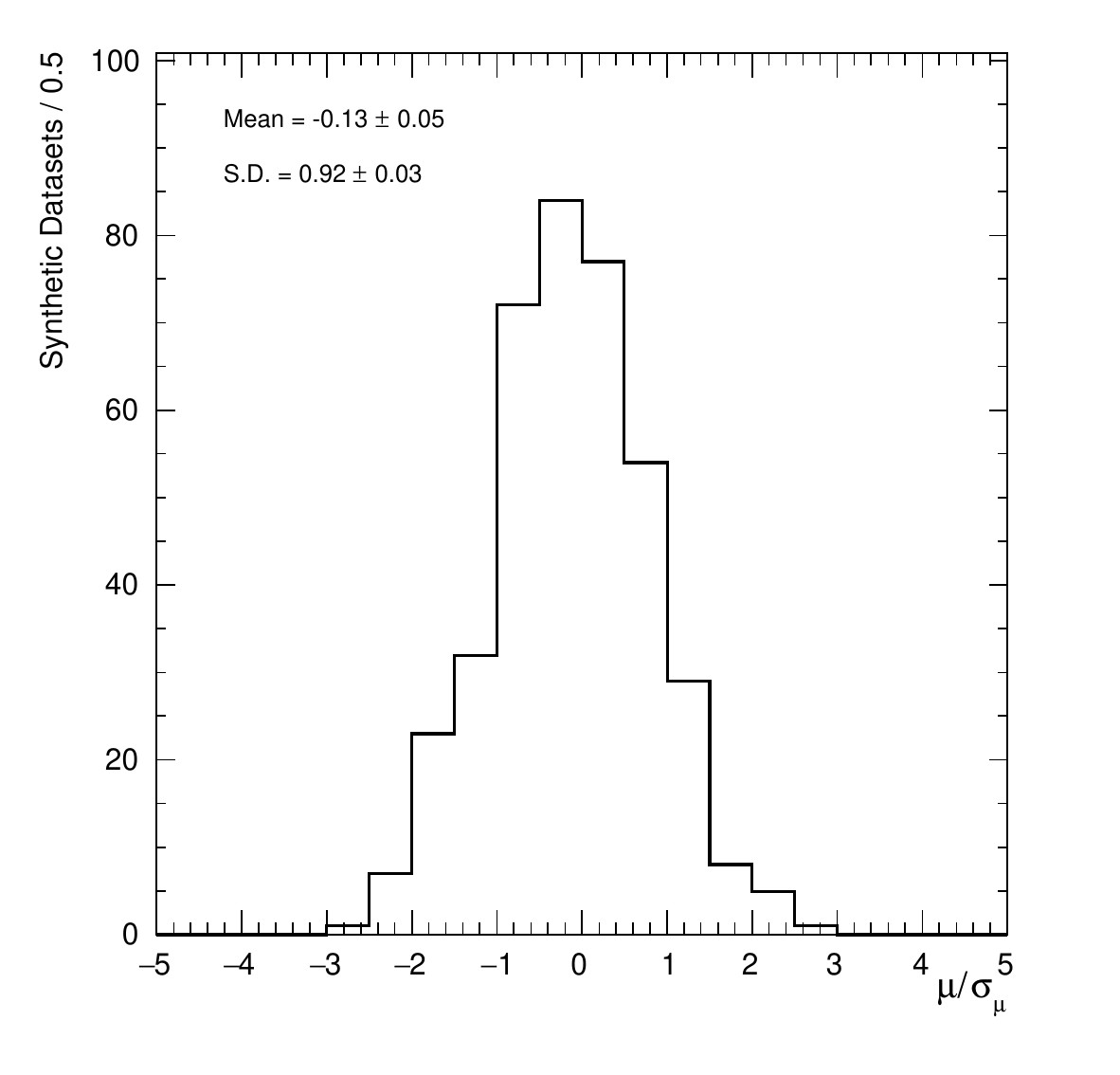}}
  \caption{Ensemble test distributions of fitted signal strength, normalised to uncertainty, for each of the two mass hypotheses considered: \subref{fig:npdd_pull_a} and \subref{fig:npdd_pull_b} signal located on kinematic peak of $m(\phi,\gamma)$ distribution, \subref{fig:npdd_pull_c} and \subref{fig:npdd_pull_d} signal located on smoothly falling region of $m(\phi,\gamma)$ distribution.
In \subref{fig:npdd_pull_a} and \subref{fig:npdd_pull_c} the background model is obtained by using once each of the $n$ events in the GR, while in \subref{fig:npdd_pull_b} and \subref{fig:npdd_pull_d} the GR is sampled $n$ times.\label{fig:npdd_pull}}
\end{figure}

The background modelling procedure was applied to each dataset in the
ensemble and the statistical analysis described in
section~\ref{NPDD_stat} was applied. Two signal masses were
considered. For the first, the signal was located on the peak of the
rising kinematic edge of the $m(\phi,\gamma)$ distribution. The second
was located in the area where the $m(\phi,\gamma)$ distribution is
smoothly falling, as in the main case study. The distribution of the
signal strength normalised to its uncertainty, also referred to as the
pull distribution\footnote{The true value of the signal strength is
zero, given that these are background-only datasets.}, for each signal
mass hypothesis is shown in figure~\ref{fig:npdd_pull}. In each case,
the pull distribution is shown for the background model obtained by
directly utilising each of the $n$ events in the GR once and for that
obtained by sampling $n$ events from the GR dataset with
replacement. In all cases, the ensemble tests demonstrate that the
method provides a background model of sufficient accuracy to avoid any
substantial bias in the statistical quantification of a potential
signal. Nevertheless, for the background models obtained by directly
utilising once each of the $n$ events in the GR, the width of the pull
distribution is substantially less that the expected value of
unity. In contrast, for the background model obtained by sampling the
GR events with replacement, the width of the pull distribution is much
closer to unity.

\begin{figure}[!h]
  \centering
  \subfigure[\label{fig:npdd_fluctuation}]{\includegraphics[width=0.45\textwidth]{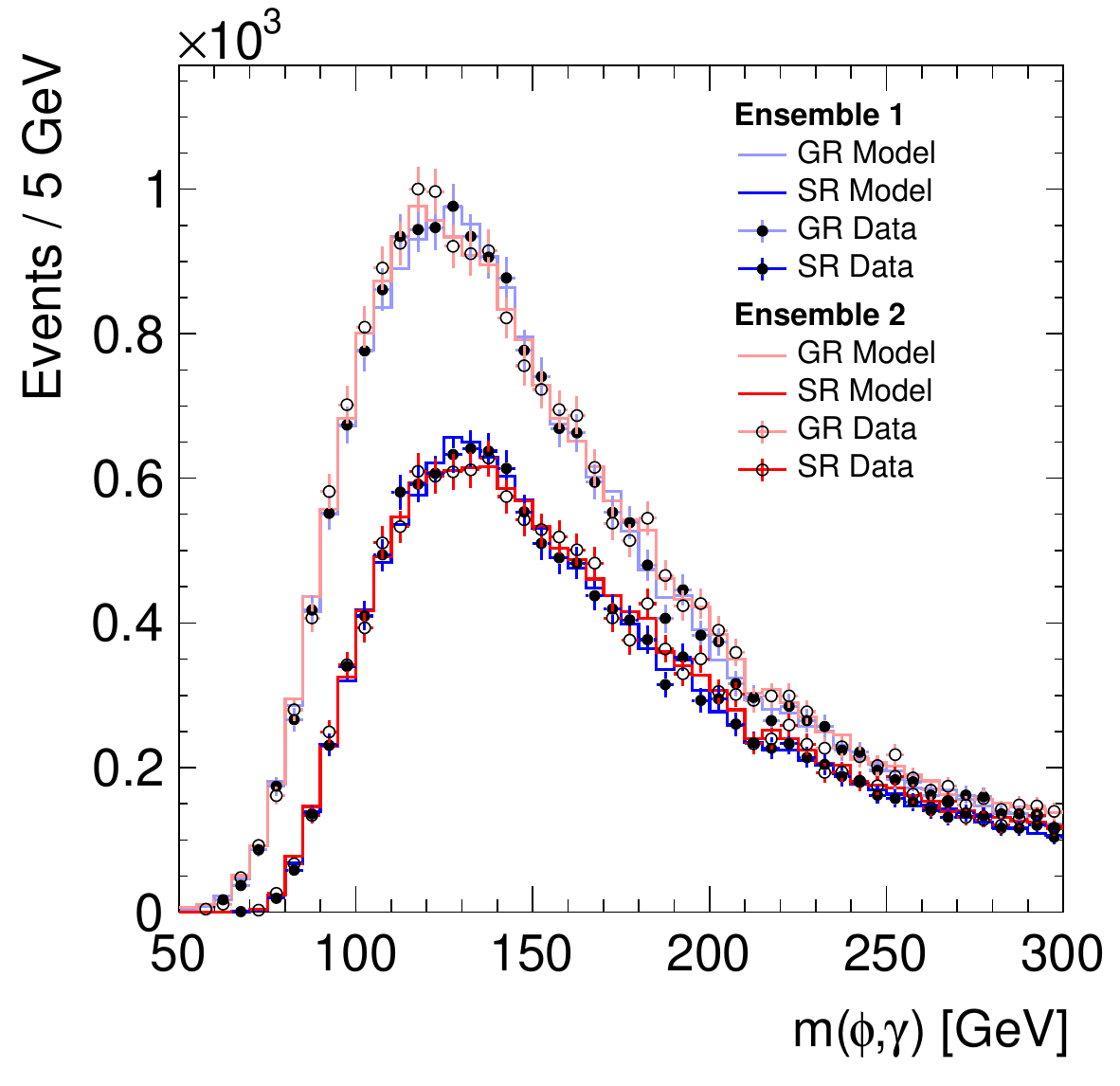}}
  \subfigure[\label{fig:npdd_pullevolution}]{\includegraphics[width=0.45\textwidth]{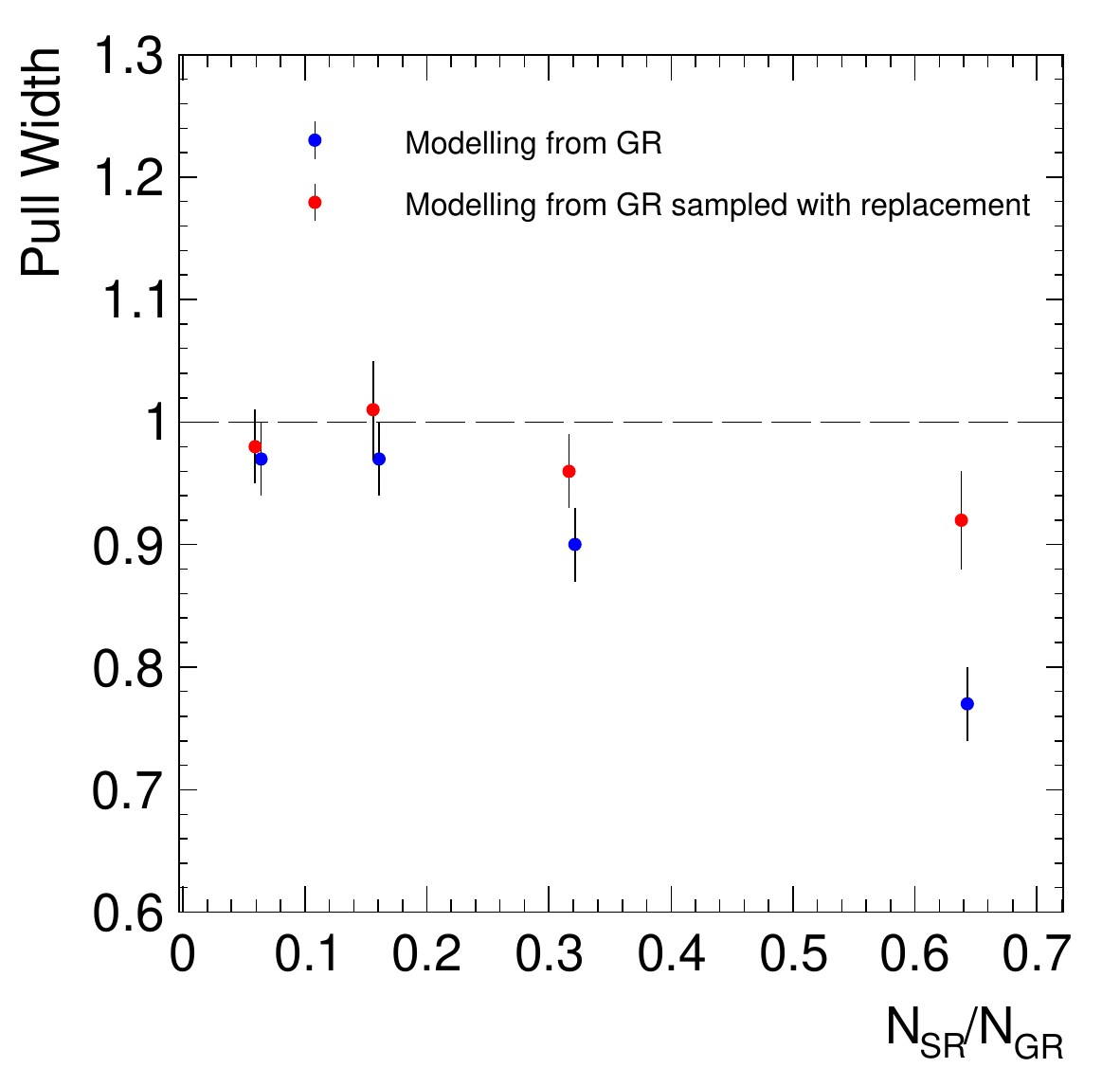}}
\caption{\subref{fig:npdd_fluctuation} Data and background model comparison for two distinct experiments in the ensemble. The background model is obtained utilising each of the $n$ GR events once. \subref{fig:npdd_pullevolution} The width of the pull distribution for ensemble tests with different SR-to-GR event fractions. The red markers have been shifted by -0.5\% along the horizontal axis for visibility.}
\end{figure}

The observed deviation in the width of the pull distribution from
unity arises because the events in the SR are not independent from the
events used to obtain the background model, since the SR is a subset
of the GR. This is illustrated in figure~\ref{fig:npdd_fluctuation}
for the case where the background model is obtained utilising each of
the $n$ events in the GR once. In this figure, the data and the
obtained background model are compared, prior to any fit, for two
distinct experiments from the ensemble. It is shown that the
background model reproduces the fluctuations in the background shape
observed in the specific experiment to a good degree. The size of this
effect scales with, $N_{SR}/N_{GR}$, the ratio of the number of events
in the SR relative to that of the GR, and is practically removed as
this ratio tends to zero. Typically, as in the case of the analyses
presented in
refs.~\cite{Aaboud:2017xnb,Aaboud:2018txb,Aaboud:2016rug,Aad:2015sda},
this ratio is of order 10\% or less. However, such low values may not
always be possible to achieve due to experimental
limitations. Nevertheless, in such cases, correct statistical coverage
can be obtained by means of a MC study.

In cases where an appropriately small value of $N_{SR}/N_{GR}$ cannot
be achieved or a MC study is not feasible, this effect is mitigated by
deriving the model by sampling $n$ events from the GR dataset with
replacement. This is clearly demonstrated in
figure~\ref{fig:npdd_pullevolution}, where the width of the pull
distribution is shown as a function of $N_{SR}/N_{GR}$ for each for
the implementations. The sampling of the GR with replacement leads to
reduced correlations between the fitted dataset and background model,
resulting in a pull distribution which is practically compatible with
unity.

\section{Background modelling with generative adversarial networks}
\begin{figure}[!h]
  \centering
  \includegraphics[width=0.85\textwidth]{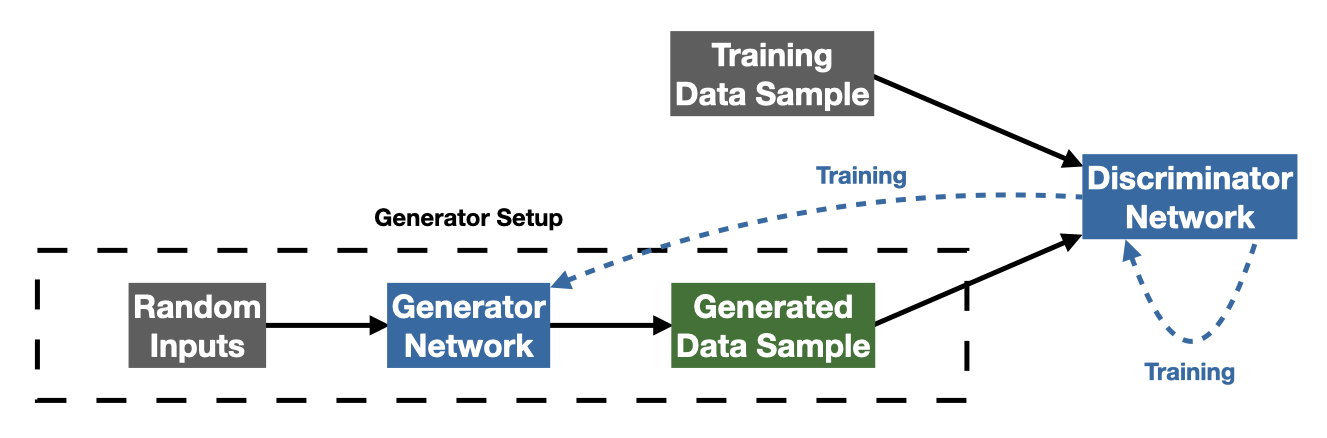}
  \caption{Schematic representation of a generative adversarial network.\label{fig:cGAN:GAN}}
\end{figure}

Generative adversarial networks (GANs)~\cite{Goodfellow:2014upx} are
generative machine learning algorithms, widely used in image
synthesis, see for example refs.~\cite{radford2015unsupervised,
  karras2017progressive, brock2018large, zhang2017stackgan, 9043519,
  YI2019101552}.  They consist of a pair of neural networks, as shown
in figure~\ref{fig:cGAN:GAN}: the generator, which is trained to learn
a generative model from a training data sample; and the discriminator,
which is simultaneously trained to discriminate the sample derived
from the generative model from the training data sample. The generator
is trained to maximise the probability that the discriminator
misclassifies the generated events for real data.  The use of GANs in
high energy physics is an active topic of research, where they can
generate events significantly faster than traditional
techniques~\cite{Alanazi:2021grv,Butter:2020tvl}, significantly
increasing the effective statistical power of a
sample~\cite{Butter:2020qhk}. This speed-up is crucial for big data
applications, for example in the HL-LHC, where large simulated
datasets are required to match the statistical precision of the data.

In the context of particle physics, where the application of machine
learning techniques is ever expanding as demonstrated by recent
reviews on the topic~\cite{Radovic:2018dip, Guest:2018yhq,
  Bourilkov:2019yoi,Schwartz:2021ftp,Karagiorgi:2021ngt,Feickert:2021ajf},
the use of GANs has been explored in applications spanning the
complete chain of event generation, simulation, and
reconstruction. The use of GANs for simulating the hard scattering
process was considered in
ref.~\cite{Otten:2019hhl,Butter:2019cae,SHiP:2019gcl}, while the use
of GANs for pileup description and detector simulation was explored in
ref.~\cite{Martinez:2019jlu} and~\cite{Ghosh:2020kkt},
respectively. Recent publications have also explored the idea of
replacing the entire reconstructed-event generation pipeline with a
GAN~\cite{Otten:2019hhl,Hashemi:2019fkn,DiSipio:2019imz,DiSipio:2019sug,Farrell:2019fsm,Alanazi:2020klf,Vallecorsa:2019ked}. Each
of these applications differ in the nature of the training data used,
but mostly use simulated training datasets. This solves the issue of
limited simulated data samples by allowing large generated samples to
be produced from much smaller datasets. However, concerns related to
simulation-based mismodelling remain, which often result in some of
the largest sources of uncertainty in searches and measurements at the
LHC.

\begin{figure}[]
  \centering
  \includegraphics[width=0.85\textwidth]{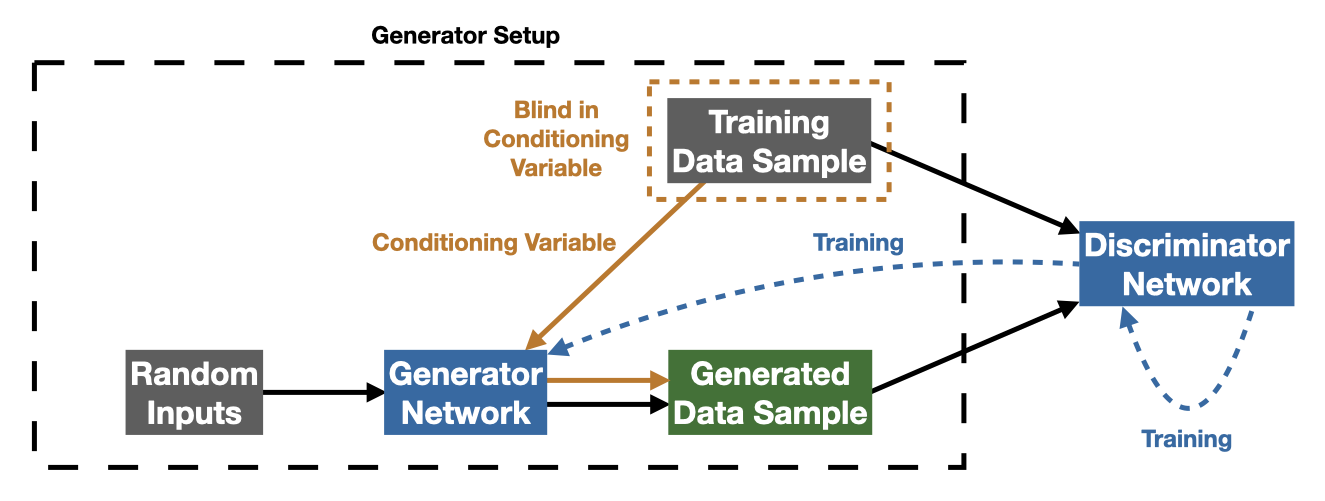}
   \caption{Schematic representation of a conditional generative adversarial network.\label{fig:cGAN:cGAN}}
\end{figure}

In our approach, this is resolved by ``blinding'' the data signal
region (SR) while training the GAN. Training a standard GAN using
blinded data explicitly, and falsely, informs the GAN that there are
no events in the SR, leading to a generative model which predicts an
absence of background events in the SR. For this reason, in this
article, a conditional GAN (cGAN)~\cite{Mirza:2014dfp}, shown in
figure~\ref{fig:cGAN:cGAN}, is used to provide a generative model of
the conditional probability distribution of the data, given the value
of the variable used to blind the dataset. To do this, it is trained
to estimate the function that maps the random latent space and the
blinding variable onto the distribution of the other feature
variables, in a similar way to how standard GANs are trained. Random
points in the latent space are then generated and provided to the cGAN
alongside the conditioning variable to estimate the other feature
variables, and then the original conditioning variable is appended to
the list of estimated feature variables. Despite being given no
information about the data in the SR, except for the values of the
conditioning variable, the predictions of the cGAN can be interpolated
or extrapolated into the SR.

To the best of the authors' knowledge, this cGAN-based technique is
the first technique of its kind to be presented. Recently,
independently and in parallel to the development of this work, a new
approach to anomaly detection in particle physics data analysis was
published that also uses a generative model of a conditional
probability distribution to model background processes in a SR through
interpolation~\cite{Hallin:2021wme}. In that case, the focus is on
anomaly detection, rather than for background modelling in a more
typical LHC analysis workflow. Furthermore, the generative model is
based on neural density estimation through a masked autoregressive
flow~\cite{papamakarios2017masked}, as opposed to GANs.

{\boldmath\subsection{Overview of case study: search for $H\to Za\to \mu\mu + {\rm jet}$}}
Searches for additional scalar or pseudo-scalar particles in the Higgs
sector are a major part of the LHC physics programme.  In particular,
the possibility for light pseudo-scalar particles, produced in the
decays of the observed Higgs boson, feature in several beyond the SM
theories~\cite{Curtin:2013fra}, including the two-Higgs-doublet model
(2HDM) and the 2HDM with an additional scalar singlet.  Searches
typically focus on Higgs boson decays into pairs of the light scalars,
or into a $Z$ boson and a light scalar.  Several searches have been
performed to date, focusing primarily either on masses of the light
resonance in excess of $4~\text{GeV}$, or considering only leptonic
decays of lighter resonances.

Recently, the ATLAS Collaboration published the first search for Higgs
boson decays to a $Z$ boson and a light hadronically decaying
resonance, $a$ ~\cite{Aad:2020hzm}. The $Z$ boson was reconstructed
from its leptonic decays to electron pairs and muon pairs
($\ell\ell$).  Masses of the light resonance $a$ between
$0.5~\text{GeV}$ and $4~\text{GeV}$ were considered and the hadronic
decay of the resonance was reconstructed inclusively as a jet. For
event selection, a multilayer perceptron
(MLP)~\cite{Goodfellow-et-al-2016} classifier is employed, which is
provided with information related to the resonance mass from a
separate MLP-based mass estimator. Finally, following a requirement on
the classifier output, an event counting approach was pursued in a
mass window of the $\ell\ell\text{j}$ system. The background
estimation was performed using a modified version of the ABCD
method~\cite{CDF:1990kbl}, where an MC-based correction was used to
account for the correlation between the $\ell\ell\text{j}$ system mass
and the classifier output.

This is a particularly interesting case study to apply this approach,
for two reasons: first, the cross-section of the main background,
$Z+\text{jets}$, is such that in the case of a ${\cal
  O}(100~\text{fb}^{-1})$ dataset, it is not feasible to generate a
simulated event sample with comparable statistical power.  Second, the
decaying resonance is identified using multi-variate methods, which
require a detailed modelling of a large number of correlations between
the relevant kinematic and jet substructure variables.  The
sensitivity of the published analysis is limited by the background
systematic uncertainties, which originate predominantly from the
insufficient size of the simulated data samples used.  By suppressing
these uncertainties, through large background samples derived directly
from the data, one may expect to first approximation a fourfold
improvement on the obtained 95\% confidence level upper limit on the
lowest light resonance masses considered, and more significant
improvements at higher masses.

In what follows, this analysis is used as a case study to implement a
cGAN-based multivariate background modelling method. The study
described here is closely aligned with the ATLAS analysis, and the
event selection is summarised below. One of the main differences with
respect to the ATLAS analysis is that for simplicity only the signal
with an $a$ mass of $0.5\,\text{GeV}$ is considered here.
Furthermore, in this study, only $Z\to\mu^{+}\mu^{-}$ decays are
considered, while the ATLAS analysis also considered $Z\to e^{+}e^{-}$
decays. This practical simplification is inconsequential for the
purposes of demonstrating the method.

\subsection{Event selection, analysis strategy and simulation}
\label{sec:cgan_event}
Events are required to contain at least two muons with transverse
momenta $p_\text{T}>5~\text{GeV}$ including an oppositely charged
pair, and a hadronic jet with transverse momentum
$p_\text{T}>20~\text{GeV}$, reconstructed using the anti-$k_t$
algorithm~\cite{Cacciari:2008gp,Cacciari:2011ma} with a distance
parameter of $0.4$. For events with more than one jet, the jet with
the highest $p_\text{T}$ is used in the analysis that follows. At
least one muon is required to have $p_\text{T}>27~\text{GeV}$ to model
the threshold imposed by the trigger, and the invariant mass of the
muons is required to be within $10\,\text{GeV}$ of the $Z$~boson mass.

This selection results in a large background arising from
$Z+\text{jets}$ events, which is mitigated using charged particle
track-based jet substructure information. Tracks with $\Delta R<0.4$
of the jet are selected if they have $p_\text{T}>0.5~\text{GeV}$ and
if their transverse and longitudinal impact parameters are compatible
with the particle being produced at the primary vertex. Events are
required to have exactly two tracks passing these requirements. Four
substructure variables are formed using these tracks: $\Delta R$
between the highest $p_\text{T}$ track and the jet axis; the ratio of
the $p_\text{T}$ of the highest $p_\text{T}$ track to the vector sum
of the track $p_\text{T}$; angularity (with weight parameter
$a=2$)~\cite{Almeida:2008yp}; and the modified correlation function
$U_1$ (with an angular exponent $\beta = 0.7$)~\cite{Moult:2016cvt}.
A neural network is used to separate signal from background events on
the basis of substructure variables described above. During the neural
network training, only signal events with tracks that originate from
the decay of the $a$ are included. A requirement is placed on the
output of the MLP which has an efficiency of 96\% and 2.5\% for $H\to
Za$ signal and $Z+\text{jets}$ background events, respectively.

A binned likelihood fit to the invariant mass distribution of the
$Z\to\mu^{+}\mu^{-}$ candidate and jet, for events passing this
selection, is used to estimate $\sigma(pp\to H)\times\mathcal{B}(H\to
Za)$. The signal and mock data are modelled using simulation, and the
background is modelled using the cGAN approach.  Inclusive Higgs boson
production in $pp$ collisions is approximated by the gluon-fusion
process alone and simulated with the Pythia 8.244 MC event generator
with the CT14nlo PDF set.  For the $H\to Za$ search, $Z+\rm{jets}$
production is expected to represent the dominant
background. Contributions such as $t\bar{t}$ production are present
only at a minor level owing to the requirement of an opposite-charge
same-flavour dilepton with an invariant mass consistent with the $Z$
boson mass. For the purposes of this study, the background from
$Z+\rm{jets}$ production alone is considered.  $Z+\rm{jets}$
production is also simulated with the Pythia 8.244 MC event generator
with the CT14nlo PDF set.  The simulated $Z+\rm{jets}$ sample contains
around $3.5\times10^8$ events and corresponds to an effective
integrated luminosity of around $160\,\text{fb}^{-1}$.

\subsection{Overview of method}
\label{sec:cganvalidationoverview}
In this method we use a GAN that is trained directly on data.  To
avoid the risk that the background model is contaminated by
potentially present signal events, the SR is ``blinded'' during the
training.  For this reason, a cGAN learns a generative model of the
data conditional probability distribution, given the value of the
blinding variable.  The reconstructed invariant mass of a signal is
typically a good choice of blinding/conditioning variable for resonant
signals where the signal is often localised to a narrow region of the
data. Nevertheless, other variables may also be used, and in fact
multiple variables may be used at once. Training a standard GAN using
blinded data explicitly, and falsely, informs the GAN that there are
no events in the SR, leading to a generative model which predicts an
absence of background events in the SR. Conversely, the cGAN learns
the distribution of the background features conditioned on the
blinding variable and so, despite being given no information about the
background in the SR, can extrapolate its prediction into the SR. The
cGAN can then be provided with the inclusive distribution of the
blinding variable for all data events, and use what it learns in the
unblinded data to interpolate the conditional generative model into
the signal region, thereby predicting the values of the other
variables. These variables could, for example, be the inputs to a
multivariate classifier. Typically, the inclusive data distribution
will often be adequate to model the inclusive distribution of the
blinding variable, as the signal contamination is small before the
application of a dedicated event selection. In that case, one may
input the raw data events directly to cGAN, or they can be used to
model the distribution, for example using a standard GAN, a histogram,
kernel density estimation, Gaussian Process regression, fitting an
analytical function, or any one of a number of other smoothing or
modelling techniques.

As a case study, the recently published search for decays of the Higgs
boson into a $Z$~boson and a light hadronic resonance is used, where a
multivariate classifier is used to target possible BSM signal
resonances. Here, the blinding variable is the invariant mass of the
$Z$~boson and the light resonance, which peaks at approximately
$125~\text{GeV}$, the mass of the observed Higgs boson. The variables
learnt by the cGAN are four jet substructure variables, largely based
on tracking information, which are used as input to a multivariate
classifier that is trained for signal to background
discrimination. Given an estimate of the inclusive invariant mass
distribution, the cGAN then provides estimates of these variables for
the background in the SR, from which the output of the classification
neural network may be calculated. A requirement on the output of the
classifier may then be applied to these generated events, providing an
estimate of the background in the SR after the multivariate
classification requirement. An extended, binned, profile likelihood
fit to the invariant mass distribution of these data events may then
be performed using a signal model and the cGAN-based background
template, to test for a possible signal.

To demonstrate the performance of the cGAN method, it is used to estimate the substructure variables for the dominant $Z+\text{jets}$ background in the $H\to Za\to \mu\mu\text{j}$ search described above. The cGANs are trained using Keras~\cite{chollet2015keras} with the Tensorflow backend~\cite{tensorflow2015-whitepaper}. Only events with exactly two tracks associated to the jet and $m_{\mu\mu\text{j}}$ in the sidebands of the $m_{\mu\mu\text{j}}$ selection of the SR, $90<m_{\mu\mu\text{j}}<123~\text{GeV}$ or $135<m_{\mu\mu\text{j}}<190~\text{GeV}$, are used in the cGAN training. Of the events that represent data in this study, 25\% are withheld from the cGAN training to quantify the performance of the cGANs. No MLP selection is applied to the events used to train the cGANs. The feature variables are transformed so that they are in the range -1 to 1.

The cGAN generator and discriminator both have architectures of 5
layers of 256 hidden nodes, each with a leaky ReLU activation function
with an alpha value of 0.4, and the input space of the generator has 9
latent dimensions and 1 additional dimension for the conditioning
variable. Other architectures were tested and were not found to
improve the performance. The generator and discriminator are both
trained using a binary cross entropy loss function, with an L2
regularisation term for their respective weights. The stochastic
gradient descent optimiser is used with Nesterov momentum, and with
gradients clipped to a value of 1. The initial network weight values
are set using He initialisation, while the biases are initialised to
0. The learning rate and its decay, in addition to the momentum, batch
size, and L2 regularisation coefficient are all determined using a
random scan over 100 sets of values. The cGANs are trained until no
significant gains are obtained.

An ensemble of five cGANs is used to reduce the uncertainty due to
imperfections in the cGAN training, which is described in
section~\ref{sec:backgroundunc}. Each cGAN is used to generate 20
times the number of data events that pass the selection, except for
the neural network and $m_{\mu\mu\text{j}}$ requirements. The cGANs
are evaluated for further training and for inclusion in the ensemble
using an analysis-specific $\chi^2$-based performance metric. This
metric is defined by comparing the data and the prediction of the
ensemble in the plane of $m_{\mu\mu\text{j}}$ and the neural network
output. Ten bins in $m_{\mu\mu\text{j}}$ for events passing the neural
network selection are used, where the total normalisation of the
ensemble in this region is set equal to the number of background
events in the region. Additionally, in the $m_{\mu\mu\text{j}}$
blinding region the five most signal-like bins in the neural network
output are used, where each neural network output region is defined to
contain 2.5\% of the total data, and the normalisation of the ensemble
in each region is corrected by the ratio of the normalisation of the
data to the ensemble of the events outside the $m_{\mu\mu\text{j}}$
blinding region. The two most signal like neural network output bins
with $123<m_{\mu\mu\text{j}}<135~\text{GeV}$ are excluded from the
definition of the metric to ensure negligible signal contamination,
and that it is uncorrelated to the signal region for events passing
the MLP selection and a possible nearby validation region. Only the
25\% of events that were withheld from the cGAN training are used in
the evaluation of this metric.

\subsection{Background modelling uncertainties}
\label{sec:backgroundunc}
Uncertainties with cGAN-based background models can arise due to the
interpolation, and imperfect performance of the cGAN.  The first
source of uncertainty is related to the size of the signal region in
the blinding variable, which is determined by the characteristics of
the signal.  This uncertainty is expected to be smaller for narrower
SRs, and for feature variables which vary more slowly across the SR.
The second source of uncertainty is related to the quality of the
training of the cGAN, which can be improved at the cost of more
analyser time or computational power, or with a larger training
dataset.

In general, these uncertainties should be quantified on a case-by-case
basis, for instance by studying the performance of the derived model
in validation regions.  A method is provided here, discussed in the
context of the case study, which is more broadly applicable.
Neglecting the uncertainty due to the interpolation, the remaining
reducible uncertainty is then estimated by comparing the predictions
of the $m_{\mu\mu\text{j}}$ distributions in the SR of the individual
cGANs to that of the ensemble. These uncertainties are correlated
between the cGANs, and so a principle component analysis is used to
obtain an orthogonal basis.

\subsection{Background model validation}
\label{sec:backgroundvalid}
\begin{figure}[]
  \centering
  \subfigure[]{\includegraphics[width=0.4\textwidth]{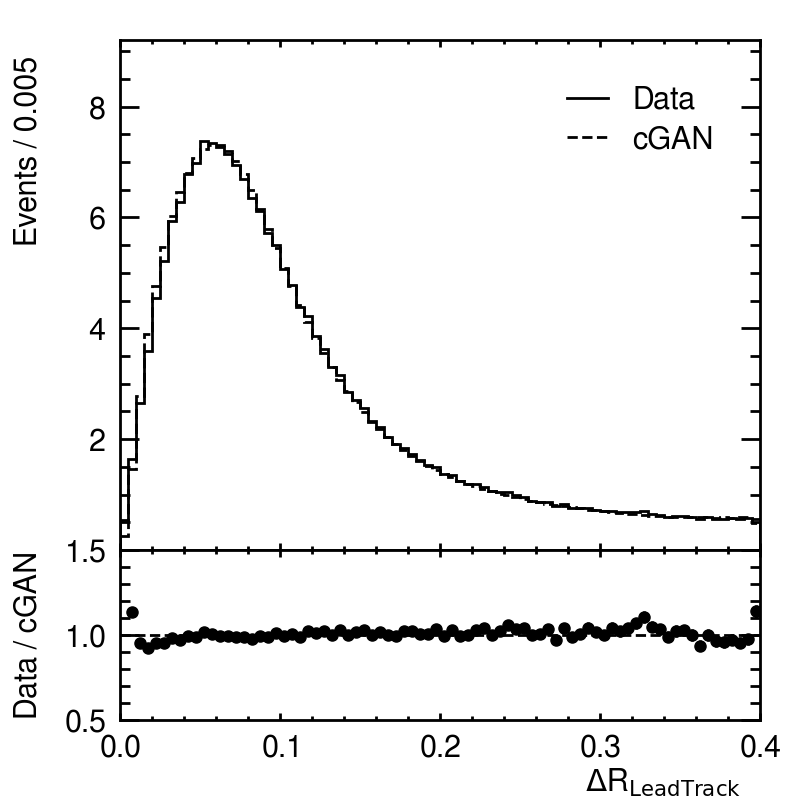}}
  \subfigure[]{\includegraphics[width=0.4\textwidth]{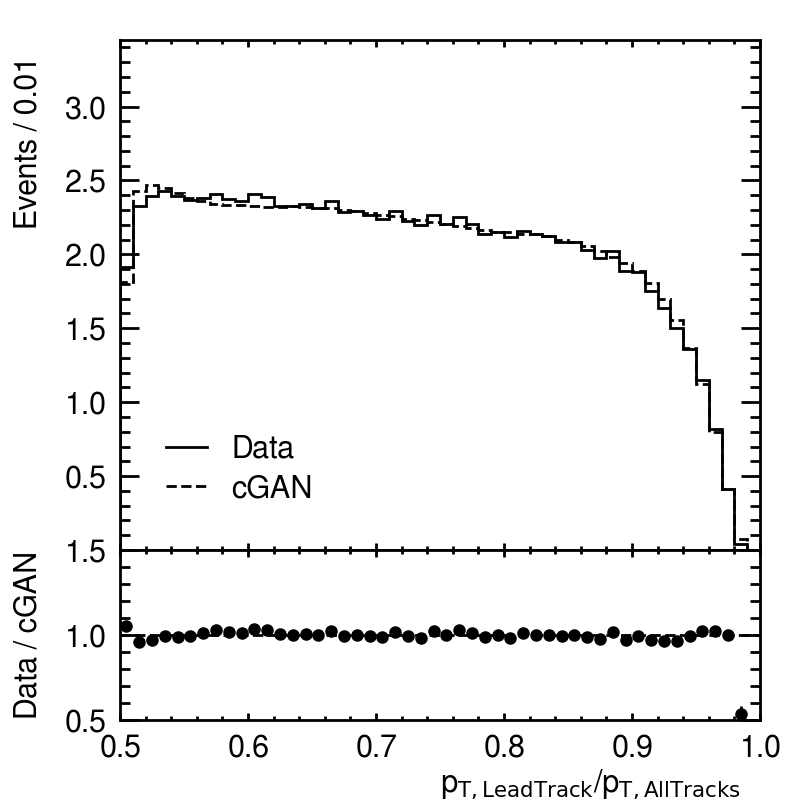}}\\
  \subfigure[]{\includegraphics[width=0.4\textwidth]{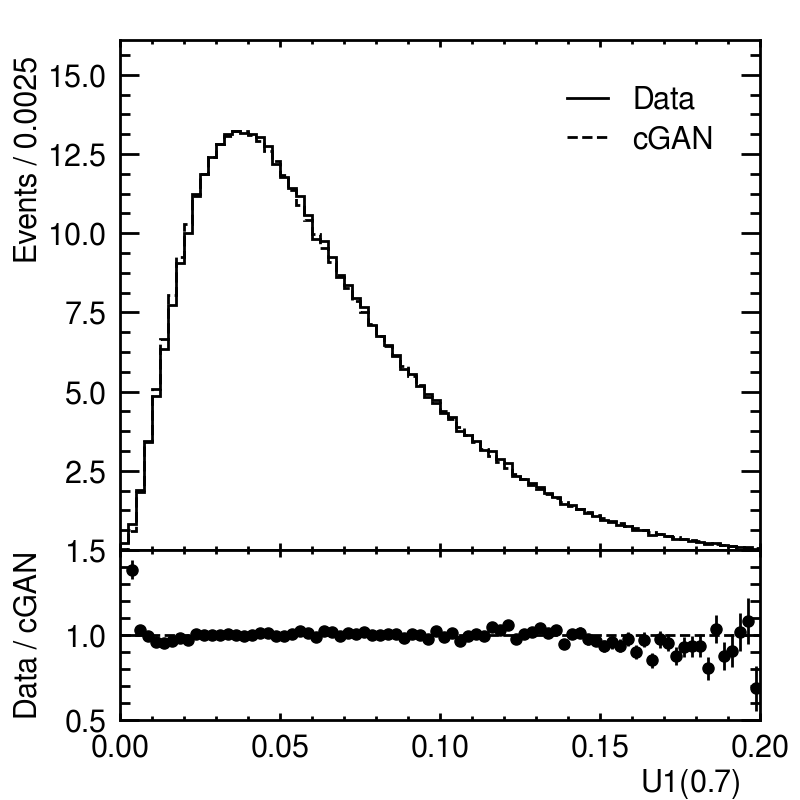}}
  \subfigure[]{\includegraphics[width=0.4\textwidth]{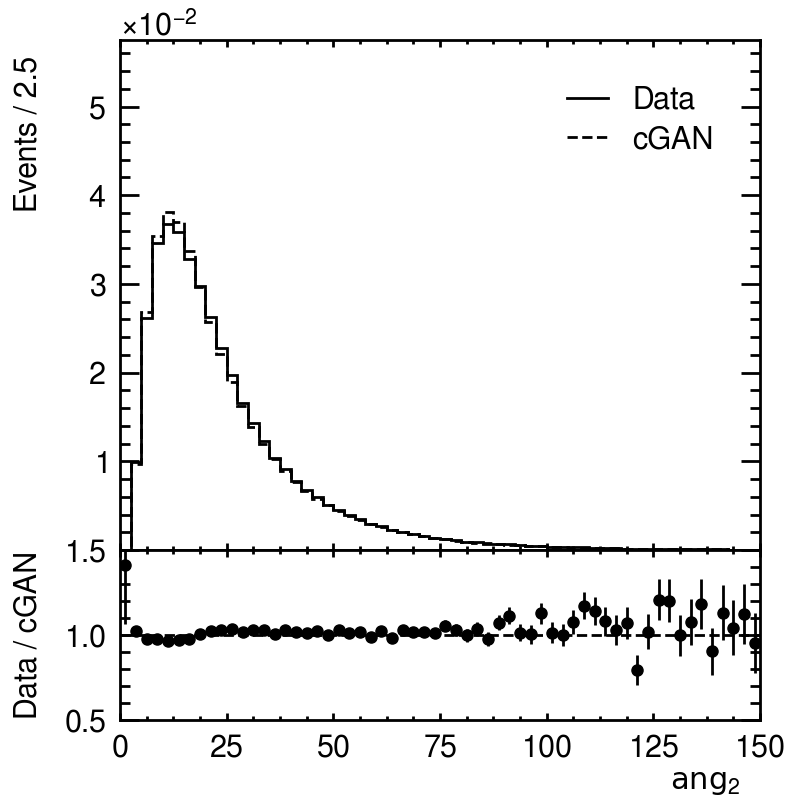}}
  \caption{Jet substructure variables in mock data and modelled by the cGAN, which are used as inputs to the classification MLP. The error bars on the markers in the lower panels represent the statistical uncertainty on the mock data.\label{fig:cGAN:substructureVars}}
\end{figure}

\begin{figure}[!htbp]
  \centering
  \subfigure[]{\includegraphics[width=0.40\textwidth]{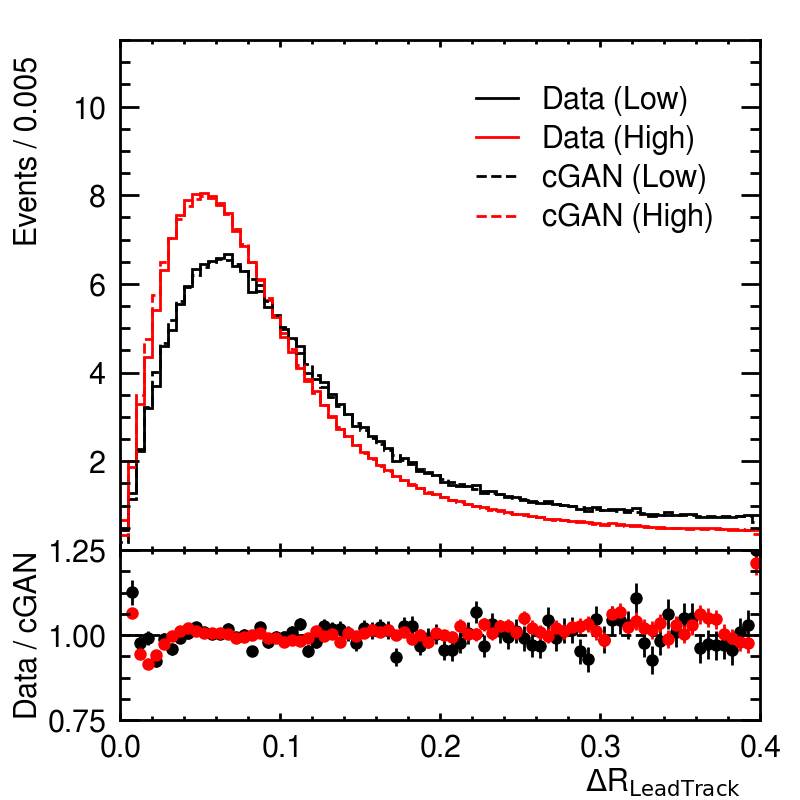}}
  \subfigure[]{\includegraphics[width=0.40\textwidth]{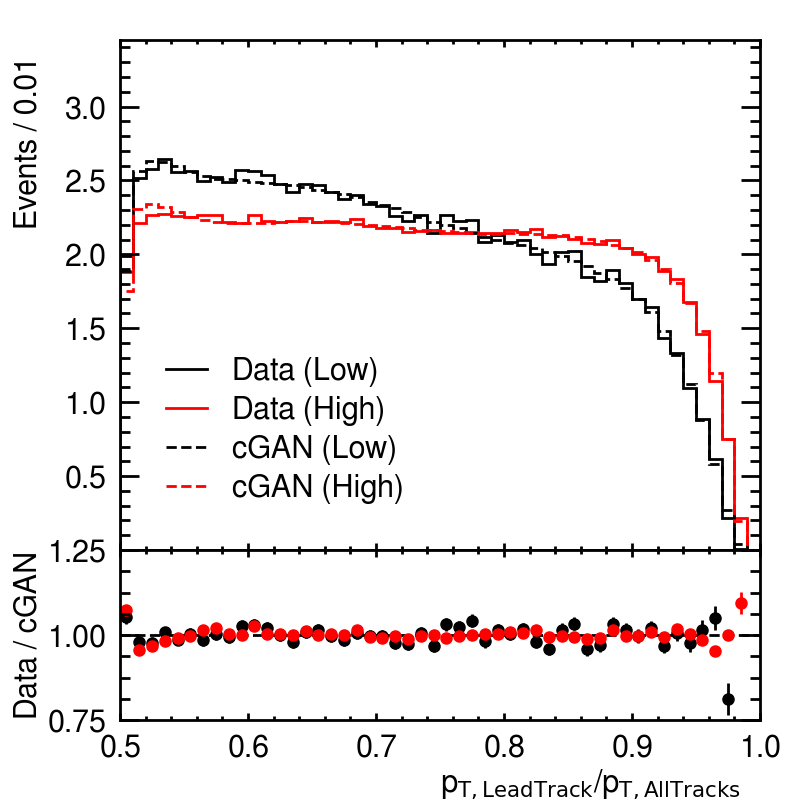}}\\
  \subfigure[]{\includegraphics[width=0.40\textwidth]{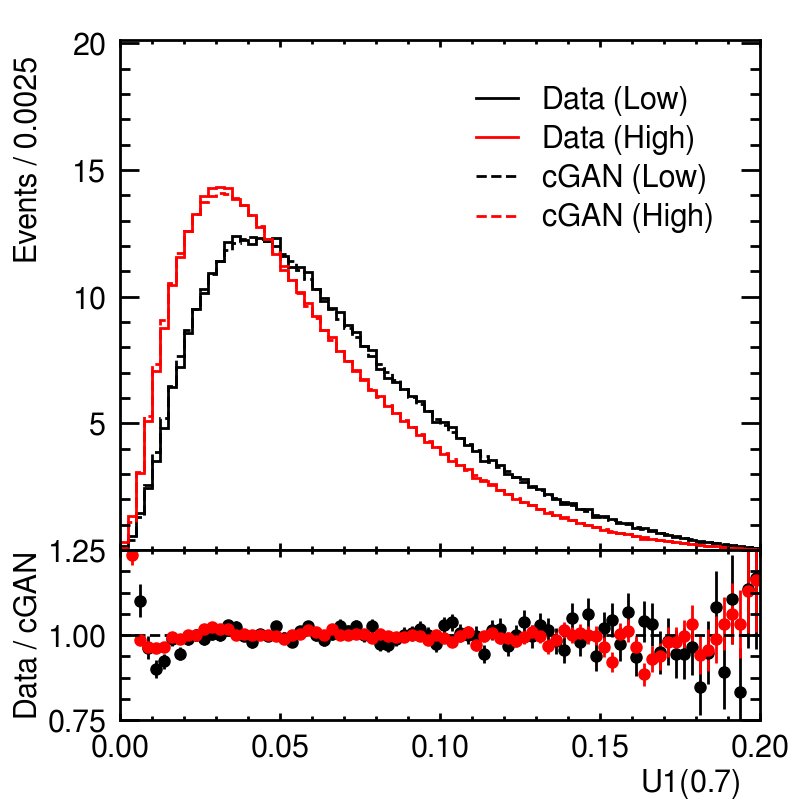}}
  \subfigure[]{\includegraphics[width=0.40\textwidth]{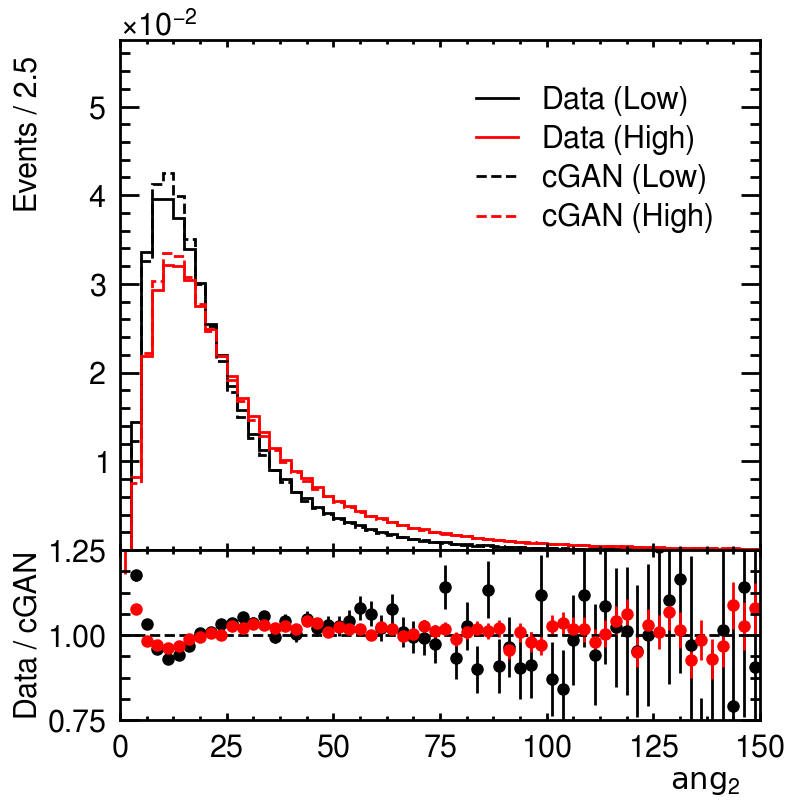}}
  \caption{Jet substructure variables in mock data and modelled by the cGAN, which are used as inputs to the classification MLP, shown for the low and high $m_{\mu\mu\text{j}}$ sidebands separately. The error bars on the markers in the lower panels represent the statistical uncertainty on the mock data.\label{fig:cGAN:substructureVarsSeparate}}
\end{figure}

The performance of the cGAN is illustrated in
figure~\ref{fig:cGAN:substructureVars}, which shows the substructure
variables in the $m_{\mu\mu\text{j}}$ SR. The cGAN is able to model
these distributions accurately, despite these data events not being
included in the cGAN training.
Figure~\ref{fig:cGAN:substructureVarsSeparate} shows the substructure
variables for the low and high $m_{\mu\mu\text{j}}$ sidebands
separately, demonstrating that the cGAN has successfully learnt the
dependence of the substructure variables on $m_{\mu\mu\text{j}}$.  The
discrepancies at the edges of the distributions are attributed to a
number of factors, including:
\begin{inparaenum}
\item the distributions often having fewer events near the edges, as compared to the cores of the distributions;
\item the edges often being sharply rising or falling; and 
\item the fact that the cGAN has fewer nearby events to learn from at the edges of the distributions.
\end{inparaenum}
Although this is not of importance in the case study presented here,
analyses that are sensitive to the modelling at these edges would
require a more thorough hyperparameter optimisation and, potentially,
a performance metric that places more importance on the quality of
modelling in these regions.

\begin{figure}[]
  \centering
  \subfigure[\label{fig:cGAN:correlation_data}]{\includegraphics[width=0.40\textwidth]{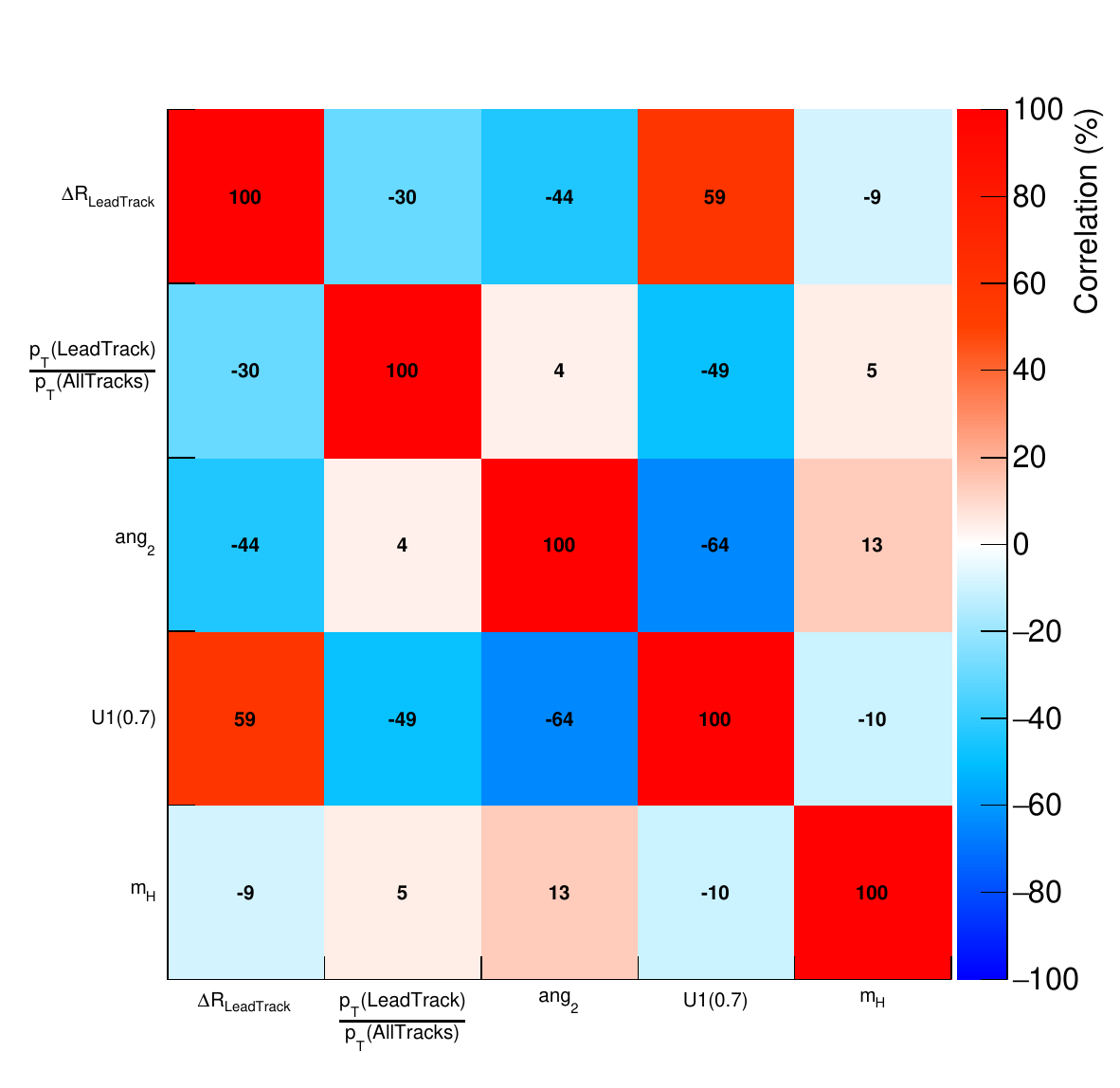}}
  \subfigure[\label{fig:cGAN:correlation_cGAN}]{\includegraphics[width=0.40\textwidth]{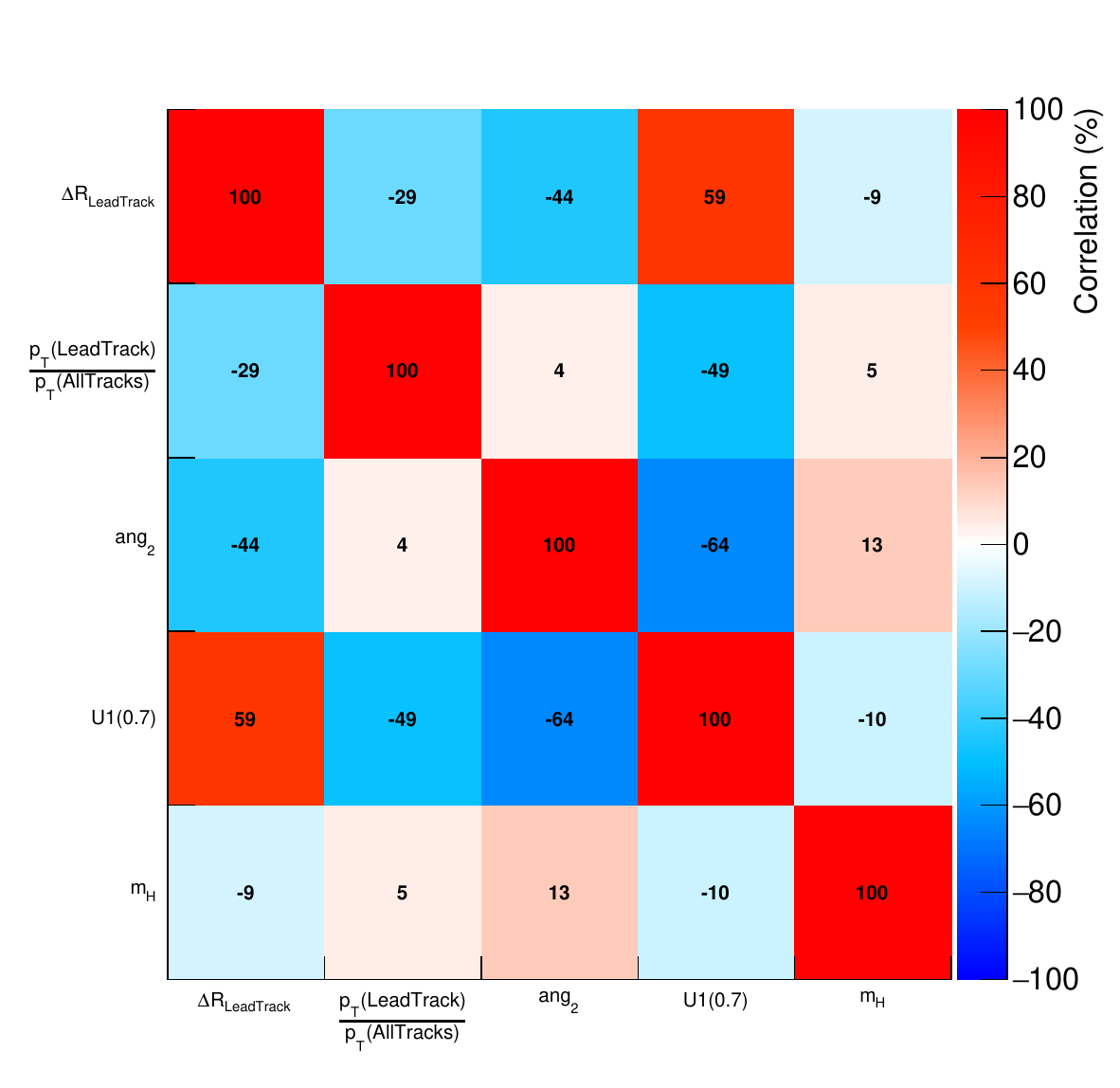}}
  \caption{Correlation matrix \subref{fig:cGAN:correlation_data} from the data and \subref{fig:cGAN:correlation_cGAN} from the cGAN.\label{fig:cGAN:correlation}}
\end{figure}

\begin{figure}[]
  \centering
  \subfigure[\label{fig:cGAN:ensemble_a}]{\includegraphics[width=0.40\textwidth]{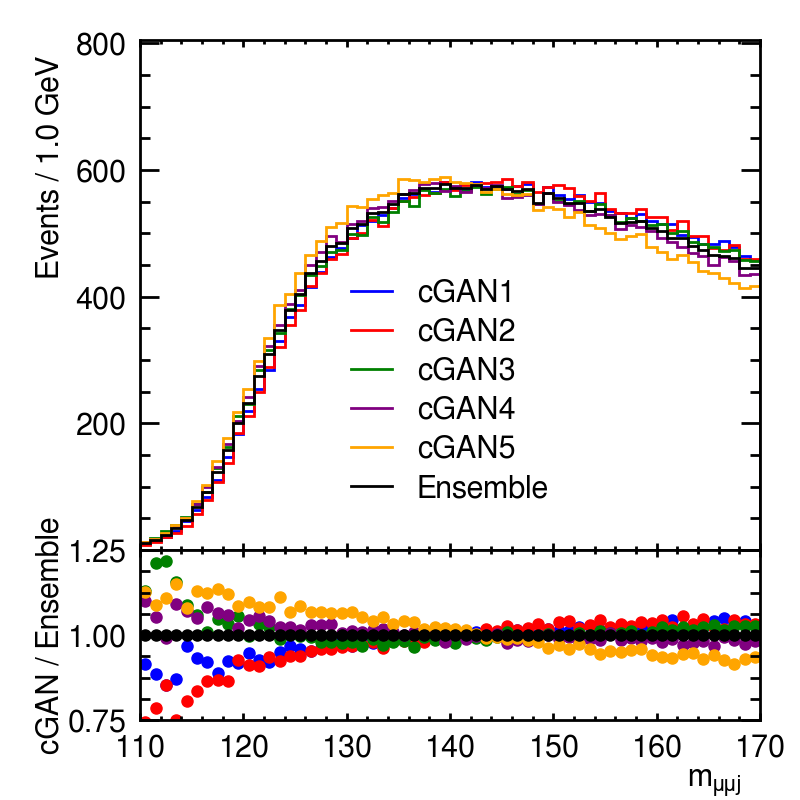}}
  \subfigure[\label{fig:cGAN:ensemble_b}]{\includegraphics[width=0.40\textwidth]{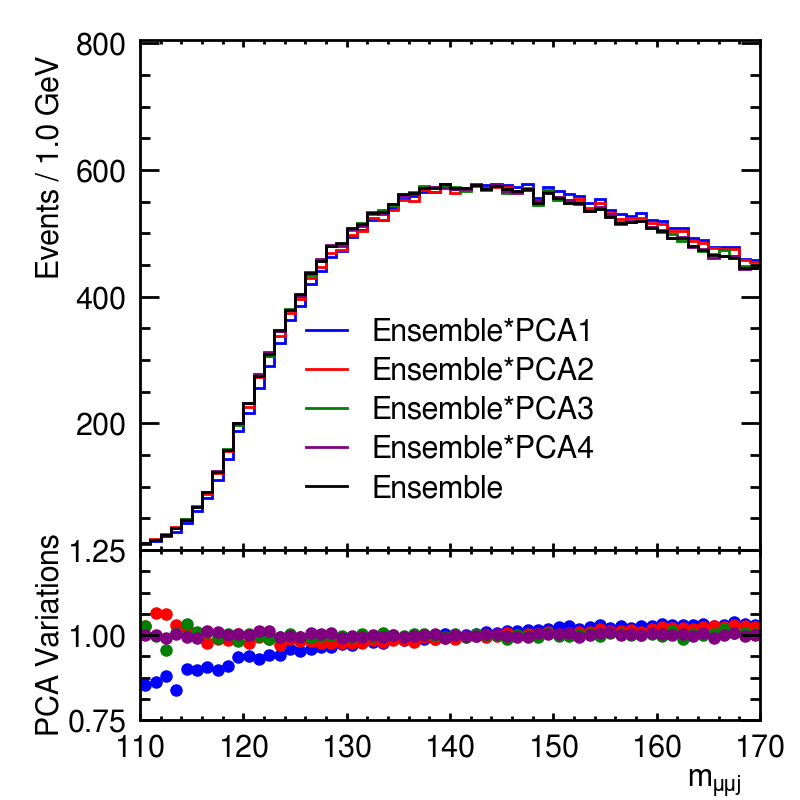}}
  \caption{\subref{fig:cGAN:ensemble_a} Variation of the top five cGANs about their ensemble, and \subref{fig:cGAN:ensemble_b} the variations that correspond to $1\sigma$ uncertainties resulting from the principle component analysis.\label{fig:cGAN:ensemble}}
\end{figure}

The obtained correlation matrix from the cGAN is compared to the data
correlation matrix in figure~\ref{fig:cGAN:correlation}. Excellent
agreement is observed. The variation of the top five cGANs about their
ensemble is shown in figure~\ref{fig:cGAN:ensemble_a}, and the
resulting uncertainties from the principle component analysis are
shown in figure~\ref{fig:cGAN:ensemble_b}.

\begin{figure}[]
  \centering
  \includegraphics[width=0.49\textwidth]{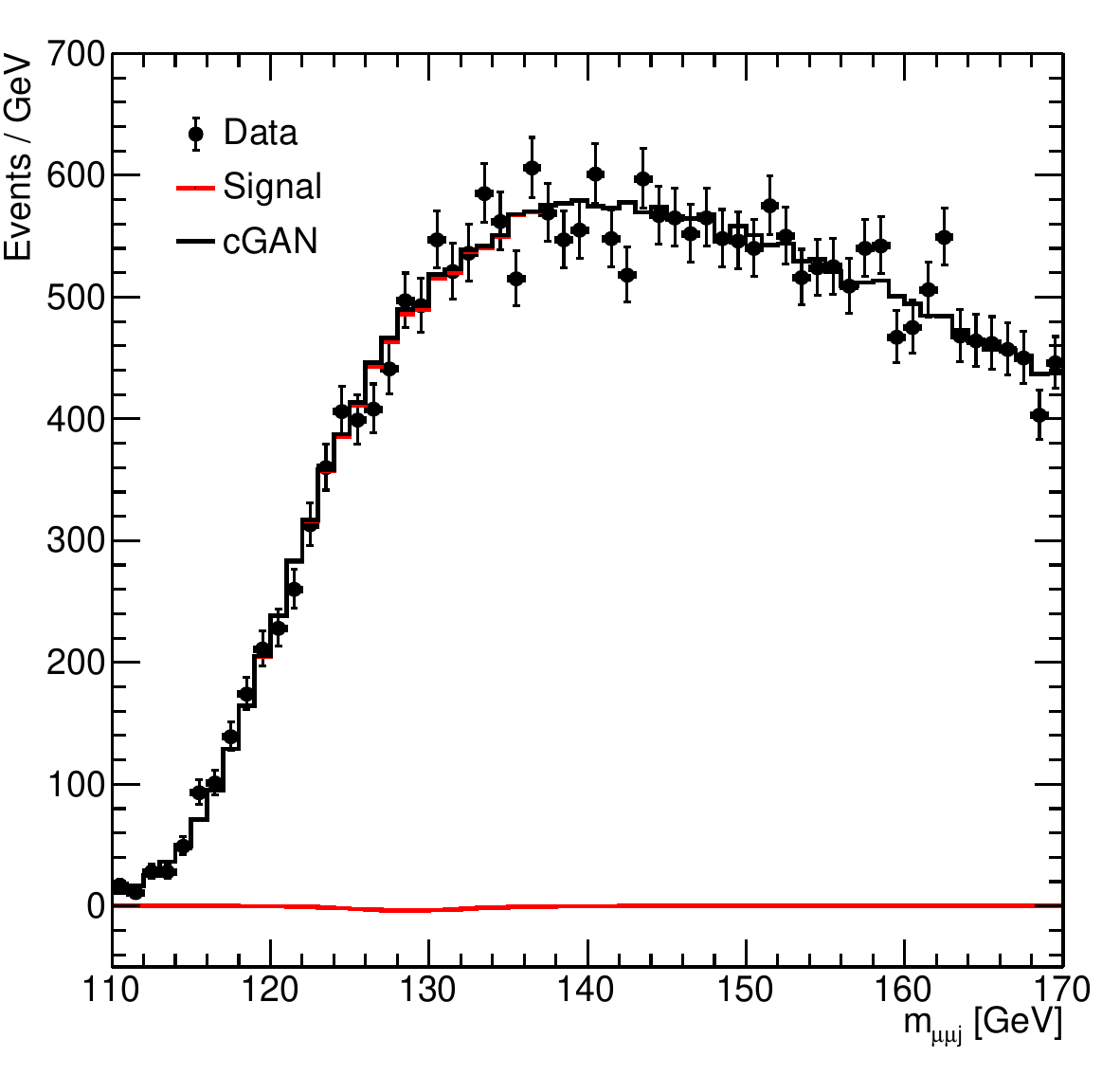}
  \caption{Post-fit $m_{\mu\mu\text{j}}$ distributions for a fit to the background-only dataset. The error bars on the markers represent the statistical uncertainty on the mock data. The sum of the signal and background models (upper red curve) and the signal model alone (lower red curve) are also shown.}
  \label{fig:cGAN:fit}
\end{figure}

An extended, binned, profile maximum likelihood fit to the
$m_{\mu\mu\text{j}}$ distribution with freely floating signal and
background normalisations is used to estimate the signal in the SR
from a fit to the background-only dataset.  The extracted signal
normalisation is $-0.003\pm 0.010$ times its predicted value, assuming
a SM Higgs boson cross
section~\cite{LHCHiggsCrossSectionWorkingGroup:2016ypw} and a
branching fraction $\mathcal{B}(H\to Za)=100\%$.  This is compatible
with the lack of signal in the dataset used in the fit.  The values of
all fitted parameters are given in table~\ref{fig:fit_gan_table},
along with their uncertainties.  The $m_{\mu\mu\text{j}}$
distributions after the likelihood fit are shown in
figure~\ref{fig:cGAN:fit}.

This case study demonstrates that the proposed background modelling
method is capable of accurately modelling a set of correlated
variables, and of interpolating over significant distances in the
conditioning variable. The performance of this method is expected to
improve in the case of smaller distances in the conditioning variable,
for example when the signal exhibits a narrow resonance in the
blinding variable.

\FloatBarrier

\begin{table}[!h]
\centering
\begin{tabular}{|c|c|c|}
\hline
Parameter & Value & Uncertainty ($\pm1\sigma$)    \\
\hline
$\mu_{\text{signal}}$   & $-0.003$    &  $\pm0.010$  \\
\hline
$\mu_{\text{bkgd}}$       & $1.001$    & $\pm0.008$          \\
\hline
Shape uncertainty 1 & $-0.36$ & $\pm0.27$ \\
\hline
Shape uncertainty 2 & $-0.31$ & $\pm0.52$ \\
\hline
\end{tabular}
\caption{Post-fit parameter values and their associated uncertainties.\label{fig:fit_gan_table}}
\end{table}

\subsection{Ensemble test using synthetic datasets}
\label{sec:cGAN:toys}
To further investigate the performance of the cGAN background
modelling strategy, ensemble tests are performed using synthetic
datasets. The results presented use 100 synthetic datasets, each
consisting of $1.5\times10^{6}$ background events generated to follow
the function
$1+x_\text{c}+2x_\text{s}+2x_\text{c}^2+x_\text{s}^2+x_\text{c}x_\text{s}$,
and $0.5\times10^{6}$ signal events generated from a Gaussian PDF in
$x_\text{c}$ with mean 0.5 and standard deviation 0.025, with
$x_\text{c}$ and $x_\text{s}$ in the range $[0, 1]$. The conditioning
variable is $x_\text{c}$ and the region $0.45<x_\text{c}<0.55$ is
blinded in the training of the cGANs. The selection $x_\text{s}>0.95$
is applied, and this requirement is relaxed in the training of the
cGANs. The average number of events in the $x_\text{c}$ blinding
region that pass the $x_\text{s}$ selection is 11000, which is about
1.9 times higher than the analogous number for the validation study
described in section~\ref{sec:cganvalidationoverview}, and if this
were used in a single-bin counting experiment these events would have
a statistical uncertainty of about 1\%. Due to these very large
synthetic datasets, this study is a highly stringent test of the
robustness of the background model.

For each synthetic dataset, 100 cGANs are trained using
$1\times10^{6}$ background events, minus the events that are blinded
in the training. The cGAN training procedure, architecture and
hyperparameter optimisation procedure largely resemble that of the
cGANs described in section~\ref{sec:cganvalidationoverview}. In
particular, the same set of hyperparameters is optimised using a
random scan over 100 sets of values, while the architecture and other
hyperparameters are the same. An ensemble is built from the five most
performant cGANs as determined with a similar $\chi^2$-based
performance metric, defined using the remaining $0.5\times10^{6}$
background events. The main exceptions to this are that the number of
output (input) nodes of the generator (discriminator) network differ,
the limits of the scan range of the number of training steps used were
lowered, and due to the associated computational cost these cGANs were
not trained until their performance approximately saturated.

\begin{figure}[!htbp]
  \centering
  \includegraphics[width=0.49\textwidth]{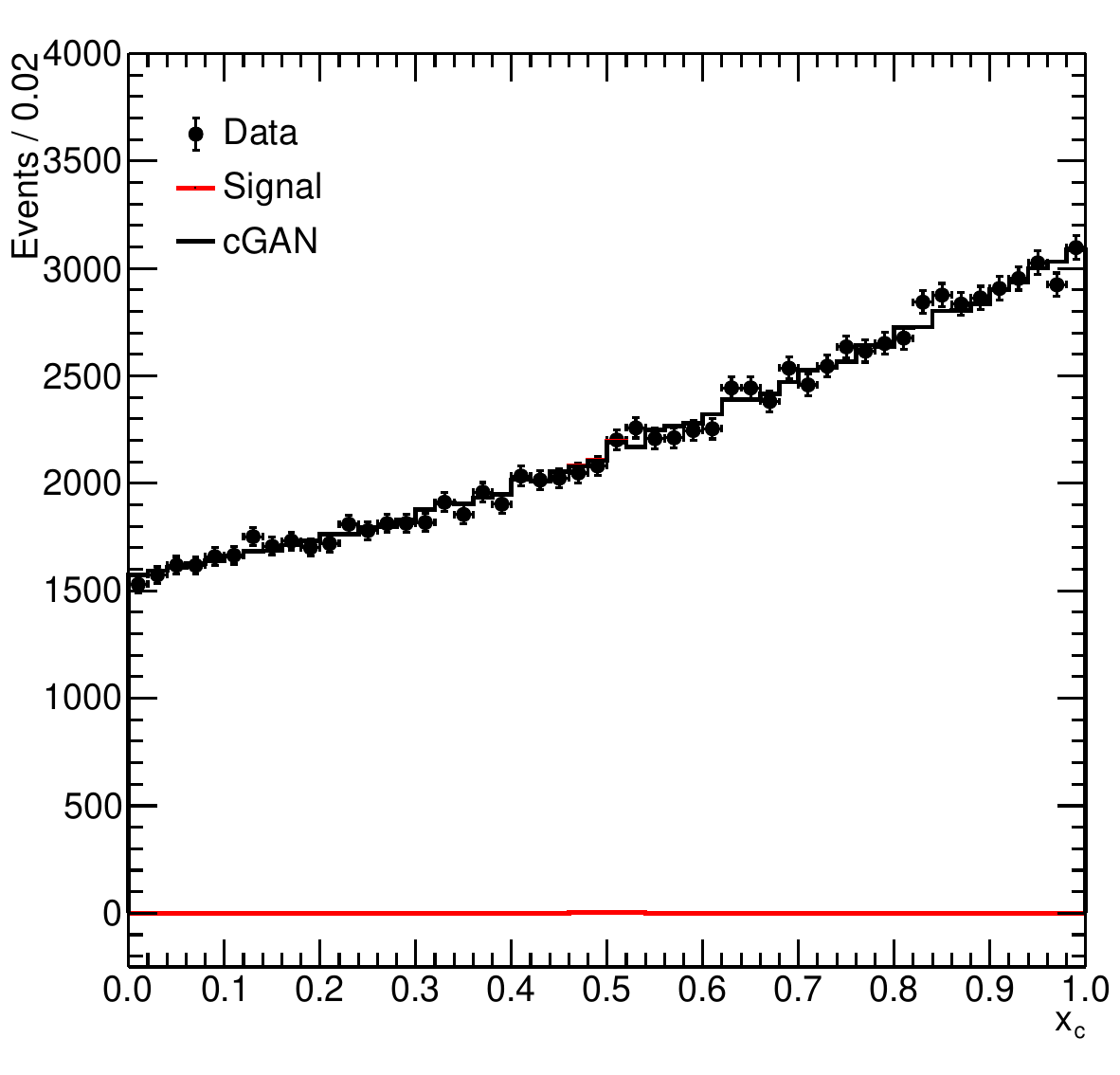}
  \caption{Post-fit $x_\text{c}$ distributions for a fit to the background-only dataset of one of the ensemble tests. The error bars on the markers represent the statistical uncertainty on the mock data. The sum of the signal and background models (upper red curve) and the signal model alone (lower red curve) are also shown.}
  \label{fig:cGAN:toyfit}
\end{figure}

Background shape uncertainties are estimated using a method similar to
that described in section~\ref{sec:backgroundunc}. An extended,
binned, profile maximum likelihood fit to the $x_\text{c}$
distribution of events passing the $x_\text{s}$ selection is performed
for each synthetic dataset, using the leading two background shape
uncertainties. The pre-fit normalisation of the cGAN-based background
estimate is set equal to the average number of synthetic data events
passing the $x_\text{s}$ selection. Figure~\ref{fig:cGAN:toyfit} shows
the post-fit $x_\text{c}$ distributions for a fit to the
background-only events passing the $x_\text{s}$ selection from one of
the synthetic datasets. This exemplifies the large number of events
generated that pass the $x_\text{s}$ selection for these synthetic
datasets, and therefore the sensitivity with which this study tests
this method.

\begin{figure}[!htbp]
  \centering
  \subfigure[\label{fig:cGAN:toys_a}]{\includegraphics[width=0.4\textwidth]{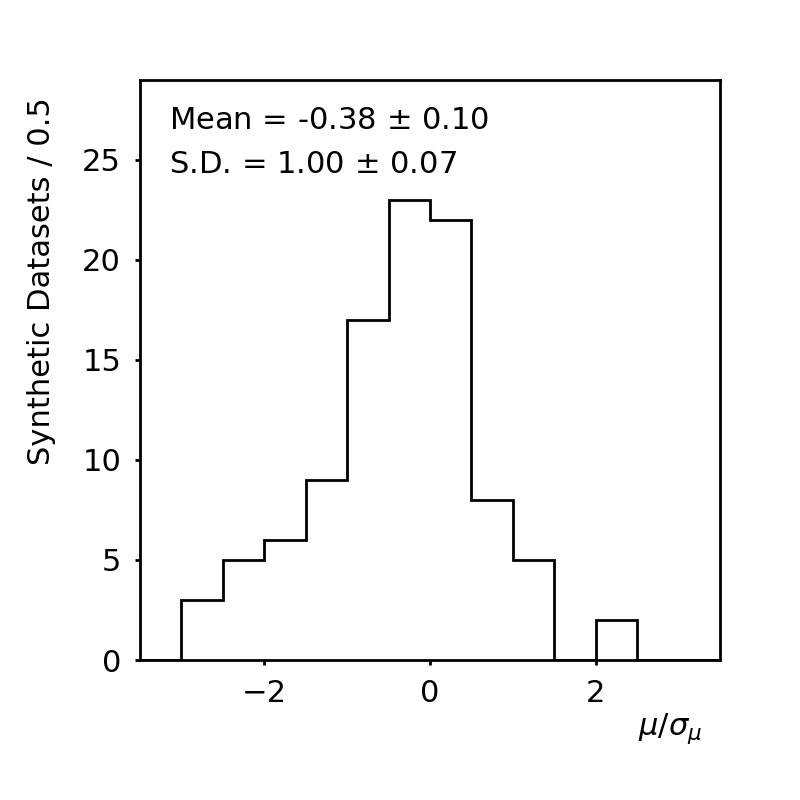}}
  \subfigure[\label{fig:cGAN:toys_b}]{\includegraphics[width=0.4\textwidth]{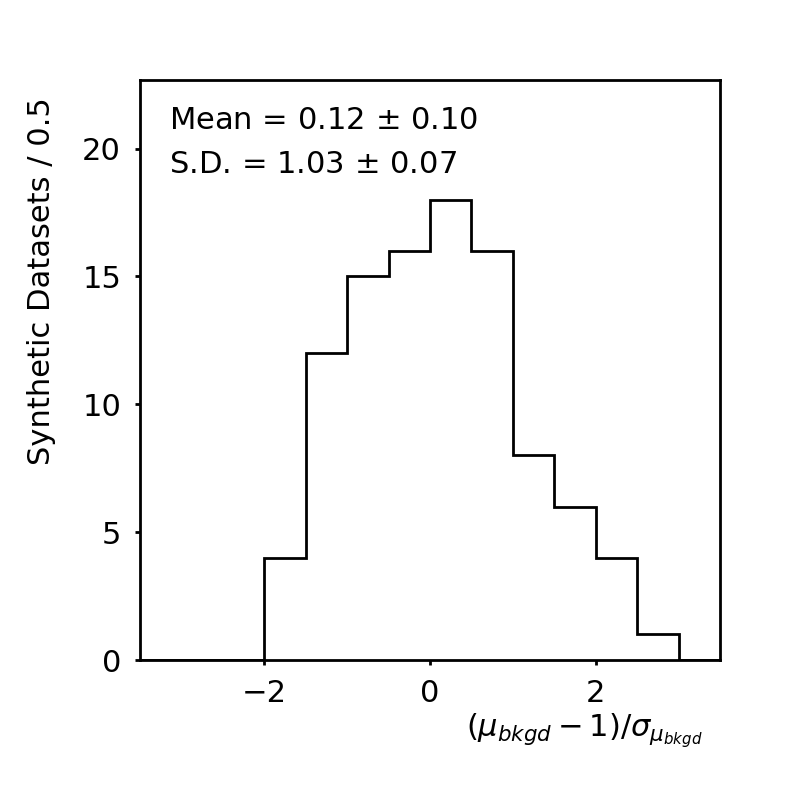}}\\
  \subfigure[\label{fig:cGAN:toys_c}]{\includegraphics[width=0.4\textwidth]{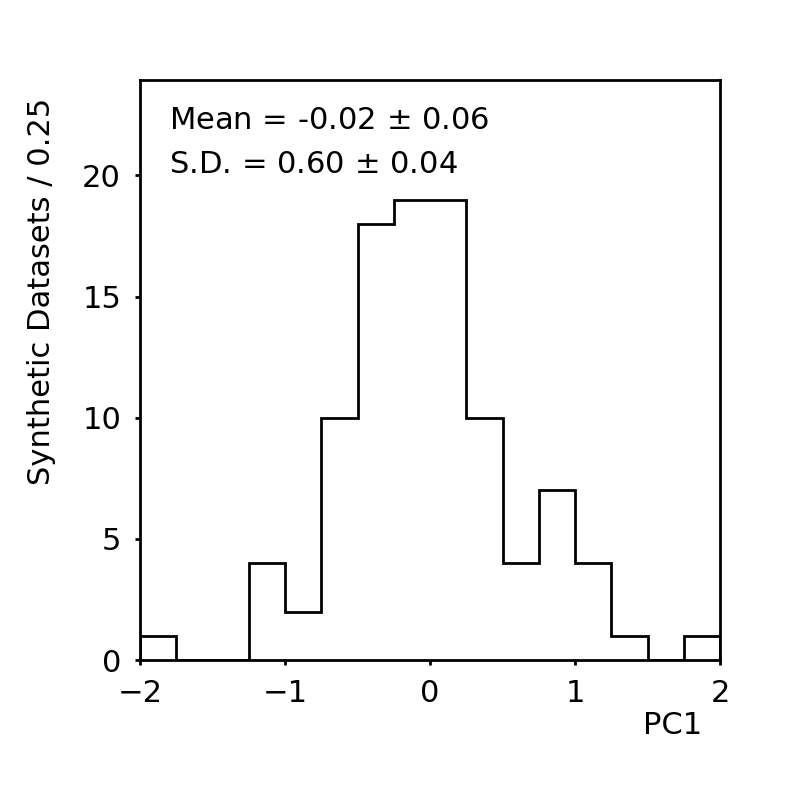}}
  \subfigure[\label{fig:cGAN:toys_d}]{\includegraphics[width=0.4\textwidth]{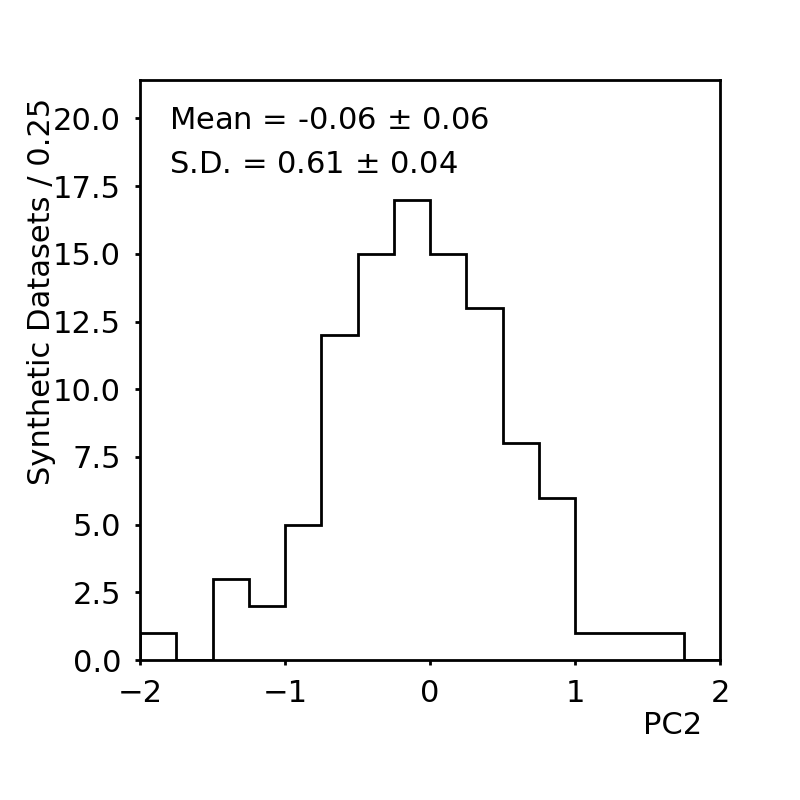}}
  \caption{The pull distribution for \subref{fig:cGAN:toys_a} the signal strength and \subref{fig:cGAN:toys_b} the background normalisation. Furthermore, the post-fit values of the nuisance parameters associated with \subref{fig:cGAN:toys_c} the leading and \subref{fig:cGAN:toys_d} the sub-leading background shape uncertainties, are also presented. These results are extracted from fits to the synthetic datasets described in section~\ref{sec:cGAN:toys}.\label{fig:cGAN:toys}}
\end{figure}

The distribution of the signal (background) normalisations extracted
from the fits (minus 1), normalised to their post-fit uncertainties,
are shown in figure~\ref{fig:cGAN:toys_a}
(figure~\ref{fig:cGAN:toys_b}). The distributions of the post-fit
values of the nuisance parameters associated with the leading
(sub-leading) background shape uncertainties are shown in
figure~\ref{fig:cGAN:toys_c} (figure~\ref{fig:cGAN:toys_d}). It is
observed that the data are able to constrain the shape nuisance
parameters in the fit. The mean of the distribution of the post-fit
signal normalisations, normalised to their post-fit uncertainties, is
-0.38. Given the standard deviation of approximately unity for this
pull variable, the mean of 100 of these quantities should have an
uncertainty of about 0.1. Thus, for this implementation of the cGAN
method, a small downward bias in the resulting estimates is
observed. This bias may originate from limitations in the training of
the cGANs. Such limitations include: the scan range of the number of
training steps used was very likely lower than optimal; and dedicated
reoptimisations of the architecture and many of the hyperparameters
were not performed.  As discussed in section~\ref{sec:backgroundunc},
another contribution to this could originate from the interpolation in
$x_\text{c}$.

An uncertainty with a magnitude of 38\% times the post-fit signal
normalisation uncertainty could be assigned to the signal
normalisation to account for this bias, which would result in around a
7\% increase to the total uncertainty. Additionally, the increase in
the average post-fit signal normalisation uncertainty due to the shape
uncertainties is 4\%. As such, this method is providing a background
estimate with a bias and a systematic uncertainty far below the
statistical uncertainty of the signal normalisation, despite the large
dataset in this search. The observed signal bias corresponds to only
0.4\% of the average number of background events in the signal region
that pass the $x_\text{s}$ selection, which is very small with respect
to the typical MC-based modelling uncertainties. Furthermore, if the
bias is due to limitations in the training of the cGANs, it is likely
possible to improve the performance of this method.

\section{Summary}
A novel and widely applicable non-parametric data-driven background
modelling method was presented, addressing typical shortcomings of
often employed strategies, such as direct simulation and parametric
models.  Two techniques were discussed for its realisation. The first
implementation uses data from a relaxed event selection to estimate a
graph of conditional probability density functions of the variables
used in the analysis. In the process the significant correlations
between the variables are accounted for. A set of
``pseudo-candidates'' is then generated through ancestral sampling
from this graph, which is subsequently propagated through the event
selection.  In the second technique, a generative adversarial network
is trained to estimate the joint probability density function of the
variables used in the analysis.  As the training is performed on a
relaxed event selection which excludes the signal region, the network
is conditioned on the chosen blinding variable.  Afterwards, the
conditional probability density function is interpolated into the
signal region to model the background.  The application of the two
methods on two benchmark analyses was presented in detail, including
their implementation in terms of statistical inference and the
derivation of associated systematic uncertainties. The performance in
ensemble tests was also discussed. As demonstrated, the methods lend
themselves to a wide spectrum of analyses, including in cases were
other often employed methods are not applicable.

\acknowledgments
The authors gratefully acknowledge valuable feedback from the
anonymous reviewers. This project has received funding from the
European Research Council (ERC) under the European Union's Horizon
2020 research and innovation programme grant agreement
714893-ExclusiveHiggs and under Marie Sk\l odowska-Curie grant
agreement 844062-LightBosons.

\bibliographystyle{JHEP}
\bibliography{bibliography}

\end{document}